\documentclass[]{pasj01}

\begin{document} 
\Received{2016/11/1}
\Accepted{2017/6/22}

\title{FOREST Unbiased Galactic plane Imaging survey with the Nobeyama 45-m telescope (FUGIN) I: Project Overview and Initial Results}

\author{Tomofumi \textsc{Umemoto}\altaffilmark{1,2*}}%
\author{Tetsuhiro \textsc{Minamidani}\altaffilmark{1,2}}

\altaffiltext{1}{Nobeyama Radio Observatory, National Astronomical Observatory of Japan (NAOJ), National Institutes of Natural Sciences (NINS), 462-2 Nobeyama, Minamimaki, Minamisaku, Nagano 384-1305, Japan}
\email{umemoto.tomofumi@nao.ac.jp}

\altaffiltext{2}{Department of Astronomical Science, School of Physical Science, SOKENDAI (The Graduate University for Advanced Studies), 2-21-1 Osawa, Mitaka, Tokyo 181-8588, Japan}

\author{Nario \textsc{Kuno}\altaffilmark{3,4}}
\altaffiltext{3}{Department of Physics, Graduate School of Pure and Applied Sciences, University of Tsukuba, 1-1-1 Ten-nodai, Tsukuba, Ibaraki  305-8577, Japan}
\altaffiltext{4}{Center for Integrated Research in Fundamental Science and Technology (CiRfSE), University of Tsukuba, Tsukuba, Ibaraki 305-8571}

\author{Shinji \textsc{Fujita}\altaffilmark{1,3,6}}

\author{Mitsuhiro \textsc{Matsuo}\altaffilmark{1,5}}
\altaffiltext{5}{Graduate Schools of Science and Engineering, Kagoshima University, 1-21-35 Korimoto,
Kagoshima, Kagoshima 890-0065, Japan}

\author{Atsushi \textsc{Nishimura}\altaffilmark{6}}
\altaffiltext{6}{Department of Astrophysics, Nagoya University, Furo-cho, Chikusa-ku, Nagoya, Aichi 464-8602, Japan}

\author{Kazufumi \textsc{Torii}\altaffilmark{1}}

\author{Tomoka \textsc{Tosaki}\altaffilmark{7}}
\altaffiltext{7}{Department of Geoscience, Joetsu University of Education, Joetsu, Niigata 943-8512, Japan}

\author{Mikito \textsc{Kohno}\altaffilmark{6}}

\author{Mika \textsc{Kuriki}\altaffilmark{3}}

\author{Yuya \textsc{Tsuda}\altaffilmark{8}}
\altaffiltext{8}{Meisei University, 2-1-1 Hodokubo, Hino, Tokyo 191-0042, Japan}

\author{Akihiko \textsc{Hirota}\altaffilmark{9}}
\altaffiltext{9}{National Astronomical Observatory of Japan (NAOJ), National Institutes of Natural Sciences (NINS), 2-21-1 Osawa, Mitaka, Tokyo 181-8588, Japan}

\author{Satoshi \textsc{Ohashi}\altaffilmark{10,17}}
\altaffiltext{10}{Department of Astronomy, The University of Tokyo, Bunkyo-ku, Tokyo 113-0033, Japan}

\author{Mitsuyoshi \textsc{Yamagishi}\altaffilmark{11}}
\altaffiltext{11}{Institute of Space and Astronautical Science, Japan Aerospace Exploration Agency, Chuo-ku, Sagamihara 252- 5210, Japan}

\author{Toshihiro \textsc{Handa}\altaffilmark{5}}
\author{Hiroyuki \textsc{Nakanishi}\altaffilmark{5,11,12}}
\altaffiltext{12}{SKA Organization, Jodrell Bank Observatory, Lower Withington, Macclesfield, Cheshire SK11 9DL, UK}

\author{Toshihiro \textsc{Omodaka}\altaffilmark{5}}
\author{Nagito \textsc{Koide}\altaffilmark{5}}

\author{Naoko \textsc{Matsumoto}\altaffilmark{9,13}}
\altaffiltext{13}{The Research Institute for Time Studies, Yamaguchi University, 1677-1 Yoshida, Yamaguchi, Yamaguchi 753-8511, Japan}

\author{Toshikazu \textsc{Onishi}\altaffilmark{14}}
\altaffiltext{14}{Department of Physical Science, Graduate School of Science, Osaka Prefecture University, 1-1 Gakuen-cho, Naka-ku, Sakai, Osaka 599-8531, Japan}
\author{Kazuki  \textsc{Tokuda}\altaffilmark{14}}

\author{Masumichi \textsc{Seta}\altaffilmark{15}}
\altaffiltext{15}{Department of Physics, School of Science and Technology, Kwansei Gakuin University, 2-1 Gakuen, Sanda, Hyogo 669-1337, Japan}

\author{Yukinori \textsc{Kobayashi}\altaffilmark{7}}

\author{Kengo \textsc{Tachihara}\altaffilmark{6}}
\author{Hidetoshi \textsc{Sano}\altaffilmark{6}}
\author{Yusuke \textsc{Hattori}\altaffilmark{6}}

\author{Sachiko \textsc{Onodera}\altaffilmark{8}}

\author{Yumiko \textsc{Oasa}\altaffilmark{16}}
\altaffiltext{16}{Faculty of Education, Saitama University, 255 Shimo-Okubo, Sakura, Saitama, Saitama 388-8570, Japan}

\author{Kazuhisa \textsc{Kamegai}\altaffilmark{9}}

\author{Masato \textsc{Tsuboi}\altaffilmark{11}}

\author{Yoshiaki \textsc{Sofue}\altaffilmark{10}}

\author{Aya E. \textsc{Higuchi}\altaffilmark{17}}
\altaffiltext{17}{The Institute of Physical and Chemical Research (RIKEN), Wako, Saitama 351-0198, Japan}

\author{James O. \textsc{Chibueze}\altaffilmark{9,18}}
\altaffiltext{18}{Department of Physics and Astronomy, Faculty of Physical Sciences, University of Nigeria, Carver Building, 1 University Road, Nsukka, Nigeria}

\author{Norikazu \textsc{Mizuno}\altaffilmark{9}}
\author{Mareki \textsc{Honma}\altaffilmark{9}}
\author{Erik \textsc{Muller}\altaffilmark{9}}
\author{Tsuyoshi \textsc{Inoue}\altaffilmark{9}}
\author{Kana \textsc{Morokuma-Matsui}\altaffilmark{9, 11}}

\author{Hiroko \textsc{Shinnaga}\altaffilmark{5}}
\author{Takeaki \textsc{Ozawa}\altaffilmark{5}}

\author{Ryo \textsc{Takahashi}\altaffilmark{14}}

\author{Satoshi \textsc{Yoshiike}\altaffilmark{6}}
\author{Jean \textsc{Costes}\altaffilmark{6}}

\author{Sho \textsc{Kuwahara}\altaffilmark{10}}





\maketitle

\begin{abstract}
The FOREST Unbiased Galactic plane Imaging survey with the Nobeyama 45-m telescope (FUGIN) project is one of the legacy projects using the new multi-beam FOREST receiver installed on the Nobeyama 45-m telescope.
This project aims to investigate the distribution, kinematics, and physical properties of both diffuse and dense molecular gas in the Galaxy at once by observing $^{12}$CO, $^{13}$CO, and C$^{18}$O $J=1-0$ lines simultaneously. 
The mapping regions are a part of the 1st quadrant ($\timeform{10D} \leq l \leq \timeform{50D}$, $|b| \leq \timeform{1D}$) and the 3rd quadrant ($\timeform{198D} \leq l \leq \timeform{236D}$,  $|b| \leq \timeform{1D}$) of the Galaxy, where spiral arms, bar structure, and the molecular gas ring are included. 
This survey achieves the highest angular resolution to date ($\sim$$\timeform{20"}$) for the Galactic plane survey in the CO $J=1-0$ lines, which makes it possible to find dense clumps located farther away than the previous surveys. 
FUGIN will provide us with an invaluable dataset for investigating the physics of the galactic interstellar medium (ISM), particularly the evolution of interstellar gas  covering galactic scale structures to the internal structures of giant molecular clouds, such as small filament/clump/core.
We present an overview of the FUGIN project, observation plan, and initial results, which reveal wide-field and detailed structures of molecular clouds, such as entangled filaments that have not been obvious in  previous surveys, and large-scale kinematics of molecular gas such as spiral arms.

\end{abstract}

\section{Introduction}

To understand the evolutionary cycle of interstellar gas, we have to know how molecular clouds are formed from diffuse atomic gas, how dense gas is formed within the molecular clouds and how stars are formed from the dense gas.
In such studies, a high spatial dynamic range covering everything from dense clumps ($\sim 1$ pc) to giant molecular clouds (GMCs) ($\sim 50$ pc) is required.
Furthermore, many GMCs have to be observed  to trace the evolution of molecular clouds comparing their environment and internal structures.
Currently, our Galaxy is the only object for which we can satisfy the requirement.

Recently, continuum surveys of the Galactic plane such as the GLIMPSE and MIPSGAL by $Spitzer$ \citep{benj03, carey09}, the Hi-GAL survey by $Herschel$ \citep{molinari10} and the all-sky survey by AKARI \citep{doi15} have been carried out at the mid- and far-infrared wavelengths. 
These surveys made a great contribution to identifying star-forming activity, such as outflow from massive protostars  \citep{cyganowski08}, new star clusters  \citep{mercer05}, H{\sc ii} regions \citep{anderson12}, bubbles associated with star formation  \citep{watson08}, and young stellar objects  \citep{toth14}.
Submillimeter continuum surveys of the Galactic plane were also conducted by some groups (the ATLASGAL survey: \cite{schuller09};  BGPS: \cite{aguirre11}; and the JCMT Plane Survey (JPS): \cite{moore15}).
These surveys revealed the detailed structure of molecular clouds, such as filaments and dense clumps (e.g., \cite{molinari10, contreras13, moore15}).
These structures are dense regions in molecular clouds, which are thought to link to the formation of stars and clusters.

Although these continuum surveys play a very important role in studies of star formation in molecular clouds as mentioned above paragraph, they lack the velocity information. 
On the other hand,  three-dimensional data obtained from observations of spectral lines provide crucial information to reveal more detailed internal structures in molecular clouds including their dynamical state and interaction.
These are very important in understanding the evolutionary process of molecular gas from diffuse gas to dense core within molecular clouds.
Since the measurement of distance of molecular clouds is essential to derive the physical quantity of the molecular clouds, it is also important that kinematic distance of the clouds can be derived from the radial velocity.
Furthermore, by knowing the distance of molecular clouds, we can investigate the relation between the properties of molecular clouds and large-scale structures of the Galaxy such as spiral arms and bar.

Surveys of the Galactic plane with spectral lines using small telescopes, such as the CfA survey in $^{12}$CO ($J=1-0$) \citep{dame01},  the AMANOGAWA Galactic plane survey in $^{12}$CO and $^{13}$CO ($J=2-1$) \citep{handa12}, NANTEN $^{12}$CO ($J=1-0$) \citep{onishi08}, and the simultaneous $^{12}$CO, $^{13}$CO, and C$^{18}$O ($J=2-1$) survey by Osaka Prefecture University 1.85m telescope \citep{onishi13, nis15}  covered a large area and mainly revealed  the distribution of molecular clouds in the Galaxy and large-scale structures of molecular clouds at $\sim3'-9'$ resolution.
Surveys with larger telescopes were also conducted.
The Galactic Ring Survey (GRS) in $^{13}$CO ($J=1-0$) was made with the Five College Radio Astronomy Observatory 14 m telescope \citep{jackson06} and \citet{rathborne09} cataloged 829 clouds and 6124 clumps using the data. 
The Three-mm Ultimate Mopra Milky Way Survey (ThrUMMS) toward the southern sky is conducted with the Mopra telescope in $^{12}$CO, $^{13}$CO, C$^{18}$O ($J=1-0$), and CN ($J=1-0$) lines \citep{barnes15}.
The High-Resolution Survey of the Galactic plane (COHRS) is $^{12}$CO ($J=3-2$) survey using the James Clerk Maxwell Telescope (JCMT) \citep{dempsey13}.
The $^{13}$CO/C$^{18}$O ($J=3-2$) Heterodyne Inner Milky Way Plane Survey (CHIMPS) was also made with JCMT \citep{rigby16}.
These multi-line observations are a powerful tool to investigate physical properties of molecular gas.

We are  conducting the simultaneous $^{12}$CO, $^{13}$CO, and C$^{18}$O ($J=1-0$) survey of the Galactic plane, FUGIN (FOREST Unbiased Galactic plane Imaging survey with the Nobeyama 45-m telescope) project, as one of the legacy projects of Nobeyama Radio Observatory  \citep{minamidani16a}.
Since usage of multi-beam receivers is essential for a wide area mapping like the Galactic plane survey, we use the multi-beam receiver FOREST (FOur-beam REceiver System on the 45-m Telescope) receiver \citep{minamidani16b} for the FUGIN project.
The main issue that we will try to address with the FUGIN data is the evolutionary process of interstellar gas in the Galaxy.
The FUGIN data will provide crucial information about the transition from atomic gas to molecular gas, formation of molecular clouds and dense gas, interaction between star-forming regions and interstellar gas, and so on.
We will also investigate the variation of physical properties and internal structures of molecular clouds in various environments, such as arm/interarm and bar, and evolutionary stage, for example, measured by star-forming activity.
The multi-transition analysis, including other data, such as COHRS and CHIMPS will enable us to determine physical properties of molecular gas in various conditions of the Galactic plane.
A non-biased mapping survey with full-sampling and uniform sensitivity like FUGIN will lead to statistical studies of clumps/clouds.

In this paper, we provide an overview of the FUGIN project and initial results.
We present the specifications for  FOREST in section 2.
Observing strategy and data reduction are explained in section 3.
Initial results are presented in section 4, and the
summary can be found in section 5.

\section{FOREST(FOur-beam REceiver System on the 45-m Telescope)} 

FOREST (FOur-beam REceiver System on the 45-m Telescope) is the four-beam, dual-polarization, sideband-separating SIS receiver newly installed on the Nobeyama 45-m Telescope. Four beams are aligned at a quadrate of 2 $\times$ 2 with $\sim$$\timeform{50"}$ grid, and the beam size of each beam is $\sim$$\timeform{14"}$ at 115 GHz. Main beam efficiencies are 0.56$\pm$0.03, 0.45$\pm$0.02, and 0.43$\pm$0.02 at 86, 110, and 115 GHz, respectively. The IF (Intermediate Frequency) bandwidth is 8 GHz (4--12 GHz) realizing simultaneous observation of $^{12}$CO, $^{13}$CO, and C$^{18}$O $J=1-0$ transitions. The system noise temperatures including atmosphere are $\sim$ 150 and 250 K at 110 and 115 GHz, respectively, and image rejection ratio is  $\sim$ 10 dB. Each IF signal is divided into two, low (4--8 GHz) and high (7--11 GHz), bands and the high band is down-converted to 4--8 GHz band using 15 GHz PLO (Phase Lock Oscillator), therefore, all output signals from the FOREST have 4--8 GHz band. The details are described in \citet{minamidani16b}.

Additional down-conversion from 4--8 GHz is done to feed 2--4 GHz signals to 3-bit and 4 Gsps digitizers, PANDA (Progressive Analog to Digital converter for Astronomy: \cite{kuno11}) and/or OCTAD-A \citep{oyama12} and these digitized signals are transmitted via optical fibers to the FX-type correlator SAM45 (Spectral Analysis Machine for the 45-m telescope: \cite{kuno11}), which is equivalent to a part of ACA Correlator \citep{kamazaki12}. The SAM45 can process 16 IF signals simultaneously, and outputs 4096 channels per IF. Frequency coverage per IF can be set 16, 31, 63, 125, 250, 500, 1000, or 2000 MHz.

\section{Overview of FUGIN Project} 

\subsection{Observing strategy}

We will make maps of the 1st quadrant ($\timeform{10D} \leq l \leq \timeform{50D}$, $|b| \leq \timeform{1D}$; 80 deg$^{2}$) and the 3rd quadrant ($\timeform{198D} \leq l \leq \timeform{236D}$,  $|b| \leq \timeform{1D}$; 76 deg$^{2}$) of the Galaxy (figure 1) in $^{12}$CO, $^{13}$CO, and C$^{18}$O $J=1-0$ lines, simultaneously, using a new multi-beam receiver FOREST installed on the Nobeyama 45-m telescope.
Our mapping areas include the spiral arms (Perseus, Sagittarius, Scutum, and Norma arms), the bar structure, and the molecular gas ring. 
We observe both the 1st quadrant and 3rd quadrant including Outer arm for comparison.

\begin{figure*}[!ht]
\begin{center}
 \includegraphics[width=16cm]{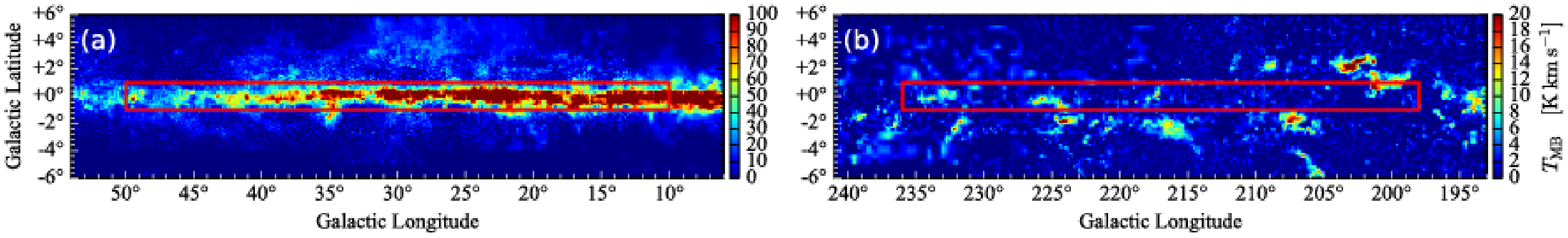}
\end{center}
\caption{Mapping regions (red boxes) for (a) the 1st quadrant ($\timeform{10D} \leq l \leq \timeform{50D}$, $|b| \leq \timeform{1D}$) and (b) the 3rd quadrant ($\timeform{198D} \leq l \leq \timeform{236D}$,  $|b| \leq \timeform{1D}$) of the Galaxy. This color map is  the CfA  survey CO map by \citet{dame01}.}
\label{fig1}
\end{figure*}

The wide coverage of gas density is one of the advantages of FUGIN.
FUGIN is the first CO Galactic plane survey at $\timeform{20"}$ resolution (the highest to date)  in the $^{12}$CO, $^{13}$CO, and C$^{18}$O $J=1-0$ lines simultaneously. 
$^{12}$CO traces total molecular gas including diffuse regions such as envelopes of molecular clouds with a density of $\sim$10$^{2}$~cm$^{-3}$, while $^{13}$CO and C$^{18}$O trace denser and optically thinner molecular gas with a density of $\sim$10$^{3}$ to 10$^{4}$~cm$^{-3}$ due to the line photon trapping  and typical optical depth (\cite{scoville74, goldreich74, scoville13}).
Therefore, we can resolve structures of both diffuse  and dense gas in molecular clouds at once.
We can also find many dense clumps without star formation which may be precursors of cluster formation with the non-biased survey in C$^{18}$O, which can be used as a tracer of dense clumps.

The FUGIN data can provide the widest spatial dynamic range ever which can cover from dense clumps ($\sim1$ pc) to structures larger than giant molecular clouds (GMCs) ($> 50$ pc).
Observations are made with the On-The-Fly (OTF) mapping mode \citep{sawada08}.
The overall map is made as a mosaic with a $\timeform{1D} \times \timeform{1D}$ sub-map.
During OTF mapping, the antenna is slewed continuously at a constant speed of $\timeform{100''}$/sec.
The data are dumped from the SAM45 spectrometer with an interval of 0.04~s.
The dumping speed corresponds to the spacing of $\timeform{4''}$ along the scan direction for the scan speed of $\timeform{100''}$/sec.
The FOREST receiver array is rotated at an angle of $\timeform{9D.46}$ with respect to the scan direction.
Two sets of scans can cover $\timeform{1D} \times \timeform{1D}$ with $\timeform{8''.5}$ spacing perpendicular to the scan direction in approximately two hours.
For each $\timeform{1D} \times \timeform{1D}$ map, scans parallel (X scan) and perpendicular (Y scan) to the Galactic plane will be made to reduce the scanning effect.
Scanning parameters are summarized in table 1.
The effective angular resolution of the final map is $\timeform{20''}$ for $^{12}$CO and $\timeform{21''}$ for $^{13}$CO and C$^{18}$O.
The grid size of the final map is $\timeform{8''.5}$.
The angular resolution is about  two times higher than the Galactic Ring Survey.
We can resolve, for example, 1 pc scale dense clumps in the main Galactic structures (arms and the bar structure) within the distance of $\sim10$ kpc and $Spitzer$ bubbles \citep{churchwell06} smaller than $\timeform{1'}$ which are the site of massive star and cluster formation.

The high sensitivity is another advantage of FUGIN.
We expect $T_{\mathrm{sys}}$ = 150 K for $^{13}$CO and C$^{18}$O, and 250 K for $^{12}$CO.
We use SAM45 with a frequency resolution of 244.14 kHz.
Considering the adopted window function, effective velocity resolution is 1.3 km~s$^{-1}$ at 115 GHz.
With the expected system temperature, we can achieve a sensitivity in $T_\mathrm{A}^{*}$ of 0.24 K for $^{12}$CO and 0.12 K  for $^{13}$CO and C$^{18}$O with the velocity resolution of 1.3 km~s$^{-1}$ by observing $\timeform{1D} \times \timeform{1D}$ region with 7.5 hours.
If the velocity resolution is reduced to 5 km~s$^{-1}$, we can achieve 0.12 K for $^{12}$CO.
We will be able to find clumps with $\sim10^2 \MO$ even at a distance of $\sim$10 kpc with this sensitivity (cf. \cite{ikeda09}).
In our target area, the integration time for the 3rd quadrant is set to be two times shorter than that for the 1st quadrant described above. This is because the major target area in the 3rd quadrant is closer than that in the 1st quadrant, and the angular resolution can be reduced for achieving a comparable spatial resolution to the 1st quadrant, and then, achieving a comparable sensitivity.
Observation parameters are summarized in table 2.

\begin{table}
  \caption{Scanning parameters. }\label{ttable1}
    \begin{tabular}{cc}
      \hline  \hline  
      RX angle & 9.$^\circ$46 \\
      Scan spacing & $8.''5$\\
      Scan length & $3600''$\\
      Scan speed & $100''$/sec \\
      Dumping time & 40 msec  \\
       Dumping spacing & $4''$\\
       Scan direction & X, Y  \\
      \hline
    \end{tabular}
\end{table}

\begin{table}
  \caption{Observation parameters. }\label{ttable2}
    \begin{tabular}{cc}
      \hline  \hline  
      Molecules & $^{12}$CO, $^{13}$CO, and C$^{18}$O {\it J}=1--0 \\
      Beam size & $14''$($^{12}$CO), $15''$($^{13}$CO and C$^{18}$O)\\
      Angular resolution & $20''$($^{12}$CO), $21''$($^{13}$CO and C$^{18}$O)\\
      Velocity resolution & 1.3 km~s$^{-1}$\\
      $T_{\mathrm{sys}}$ & 250~K($^{12}$CO), 150~K($^{13}$CO and C$^{18}$O) \\
       Expected rms ($T_\mathrm{A}^{*}$) & 0.24~K($^{12}$CO), 0.12~K($^{13}$CO and C$^{18}$O)\\
       Galactic longitue  &10$^\circ$  to 50$^\circ$, 198$^\circ$ to 236$^\circ$\\
       Galactic latitude & --1$^\circ$  to 1$^\circ$ \\
       Survey area  & 80~deg$^{2}$, 76~deg$^{2}$\\
        Angular sampling & $8.''5$ \\

      \hline
    \end{tabular}
 \end{table}

Observations are conducted according to the following procedure.
The telescope pointing is checked about every hour by observing SiO maser sources with the 40 GHz receiver H40.
The chopper-wheel method is used to correct for atmospheric and antenna ohmic losses and to get the antenna temperature, $T_\mathrm{A}^{*}$.
The intensity scale is converted from $T_\mathrm{A}^{*}$ into the main beam temperature, $T_{\mathrm{mb}}$, using the main beam efficiency of 0.43 for $^{12}$CO and 0.45 for $^{13}$CO and C$^{18}$O.
During each observing session, we observe a standard source such as W51 and Orion~KL to check the performance of the telescope once at least. 
Based on the observations of standard sources during each observing session in 2014, the intensity variations are less than 10-20\%, 10\%, and 10\% for $^{12}$CO, $^{13}$CO, and C$^{18}$O, respectively.
The calibrated $T_{\mathrm{mb}}$ scale intensity maps are consistent with those of CfA and NANTEN for $^{12}$CO, and FCRAO Galactic Ring Survey for $^{13}$CO.
Figure 2a shows a correlation plot of $T_{\mathrm{mb}}$ of $^{13}$CO $J=1-0$  for all valid ($>$ 5 $\sigma$) voxels ($l, b, v$) between Nobeyama 45-m/FUGIN and FCRAO/GRS by \citet{jackson06}. Nobeyama 45-m/FUGIN data are convolved to the beam size of FCRAO/GRS data ($\timeform{46''}$). The compared area is $l =\timeform{44D}$ to  $\timeform{50D}$ and $b = \timeform{-1D}$ to $\timeform{1D}$, and compared voxels are where both convolved FUGIN data and GRS data are above 5 $\sigma$ levels.  They are well fitted by a linear function of $T_{\mathrm{mb}}$(FUGIN) = (0.9783 $\pm$0.0005) $\times$ $T_{\mathrm{mb}}$(GRS). The 73 \% of the valid voxels are included within a range of  $\pm$~20 \% differences from the best fit.  Figure 2b shows the histogram of the fractional difference between the valid voxel values, [$T_{\mathrm{mb}}$(GRS) - $T_{\mathrm{mb}}$(FUGIN)] / $T_{\mathrm{mb}}$(FUGIN). This distribution is fitted by a Gaussian distribution with its offset of 0.006 $\pm$0.003 and standard deviation of 0.163 $\pm$0.003.  Therefore, most of the FUGIN data is consistent with the GRS data within $\sim$20\%.

\begin{figure*}[!ht]
\begin{center}
\includegraphics [width=16cm]{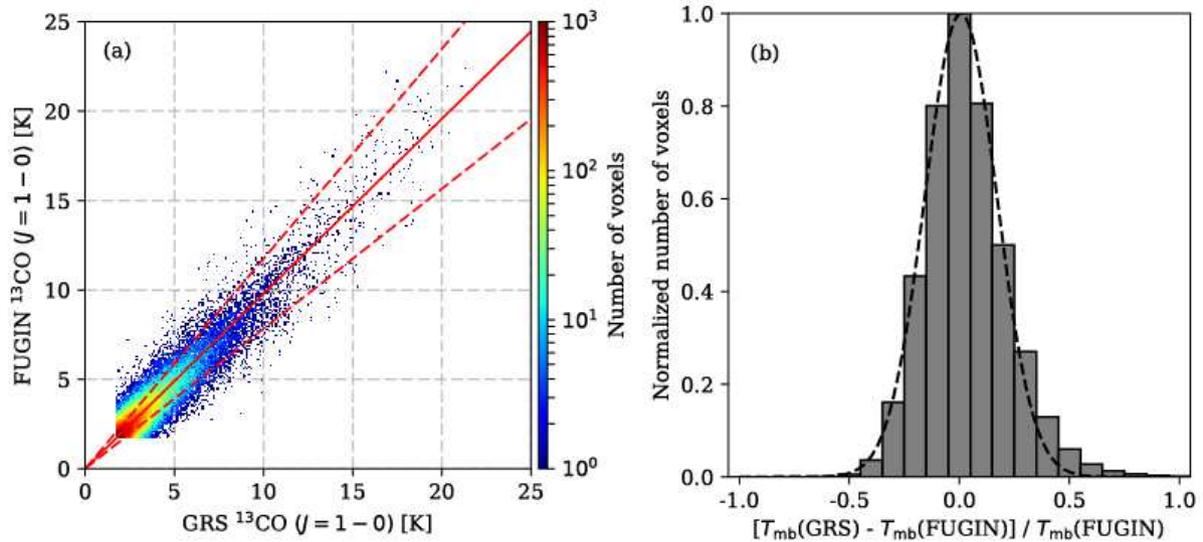}
\end{center}
\caption {Plots for comparing  $^{13}$CO $J=$1--0  data of Nobeyama 45-m/FUGIN and FCRAO/GRS. The compared area is $l =\timeform{44D}$ to $\timeform{50D}$ and $b = \timeform{-1D}$ to $\timeform{1D}$. Nobeyama 45-m/FUGIN data are convolved to the beam size of FCRAO/GRS data ($\timeform{46''}$).   All voxels  ($l, b, v$)  above 5 $\sigma$  noise level are used. (a) Correlation plot of $T_{\mathrm{mb}}$  of each valid voxel. Color scale indicates the number of voxels within each pixel (0.1 K $\times$ 0.1 K). The red line shows the best fit linear function of $T_{\mathrm{mb}}$(FUGIN) = (0.9783 $\pm$0.0005) $\times$ $T_{\mathrm{mb}}$(GRS). The red dashed lines show the $\pm$20\% differences from the best fit. (b) Histogram of the fractional difference between the valid voxel values. This histogram is normalized to the most frequent bin. The dashed curve shows the best fit Gaussian distribution with its offset of 0.006 $\pm$0.003 and standard deviation of 0.163 $\pm$0.003.}
\label{fig2}
\end{figure*}

\subsection{Data reduction} 

Data processing is performed using the NOSTAR software provided by the Nobeyama Radio Observatory combined with python scripts for pipeline reduction. 
For each $\timeform{1D} \times \timeform{1D}$ sub-map of each transition, the following data processing is performed: 
(1) Split data to each array. 
(2) Subtract baselines with a first-order polynomial function. Baseline ranges are $-200$ to $-50$ and 200 to 350 km~s$^{-1}$  for the 1st quadrant and $-100$ to $-50$ and 150 to 200 km~s$^{-1}$ for the 3rd quadrant. 
(3) Scale intensity based on the observations of standard sources in order to reduce the variation between arrays.

Maps are produced as a single FITS cube by gridding baseline-subtracted and scaled data over the desired region. The Bessel $\times$ Gaussian function is used as the convolution function, and data are gridded to $\timeform{8''.5}$ and 0.65 km~s$^{-1}$ for the 1st quadrant and $\timeform{15"}$ and 0.65 km~s$^{-1}$ for the 3rd quadrant. Velocity coverages are $-50$ to 200 km~s$^{-1}$  for the 1st quadrant and $-50$ to 150 km~s$^{-1}$ for the 3rd quadrant, respectively.  In a part of C$^{18}$O data taken in Season 2013--2014, we found strong spurious signals from AD convertor near C$^{18}$O emission, and they are eliminated and interpolated by using adjacent velocity channel data.

Additional baseline subtraction is done with the third-order polynomial function toward FITS files. Baseline ranges are automatically defined as follows: 
(1) Subtract offset and gradient by using the outermost $\sim$40 velocity channels on both sides. 
(2) Average spatially adjacent 5 $\times$ 5 spectra. 
(3) Smooth each spectrum over five velocity channels. 
(4) Regards velocity ranges of less than five sigma level of averaged and smoothed data as baseline parts. 

The final data cube is produced by scaling from the antenna temperature ($T_\mathrm{A}^{*}$ ) to main-beam temperature ($T_{\mathrm{mb}}$) scales with main-beam efficiencies.

\subsection{Policy of data release}
Based on our processed data, we will make catalogs of molecular clouds, clumps and filaments in the Galactic plane by using cloud/clump finding tools, e.g., Clumpfind \citep{williams94} and Dedrograms \citep{rosolowsky08}, filament finding tool (DisPerSE: \cite{sousbie11}),  and also catalogs of shell and arc structures that correspond to {\it Spitzer} bubbles and SNRs.

Our policy of data release is that the processed data and catalogs of clouds/clumps/filaments/shells will become open on the network after 12 months from the end of the observations of this project (scheduled for the end of May, 2017) to maximize the utility of the data for the community.
All maps and catalogs of clouds/clumps  surveyed in the 1st quadrant  (80 deg$^{2}$) and the 3rd quadrant (76 deg$^{2}$) will be presented in an upcoming paper.

\section{Initial Results}

In this section, we present a part of the data obtained from March 2014 to May 2015 as the initial results of the FUGIN project.
In this observing term, we have covered 49 deg$^{2}$  area (31\% of the planed area) in the 1st quadrant (32 deg$^{2}$ out of 80 deg$^{2}$) and the 3rd quadrant (17 deg$^{2}$ out of 76 deg$^{2}$).
In the 1st quadrant region, we have covered two regions named region A ($l= \timeform{12D}$ to $\timeform{22D}$, $b =  \timeform{-1D}$ to $\timeform{1D}$) and region B ($l= \timeform{44D}$ to $\timeform{50D}$, $b =  \timeform{-1D}$ to $\timeform{1D}$). 
In the 3rd quadrant region, we have covered three regions named  region C1 ($l= \timeform{198D}$ to $\timeform{202D}$, $b =  \timeform{-1D}$ to $\timeform{1D}$), 
C2 ($l= \timeform{213D}$ to $\timeform{218D}$, $b =  \timeform{0D}$ to $\timeform{1D}$) 
and C3 ($l= \timeform{221D}$ to $\timeform{227D}$, $b = \timeform{-1D}$ to $\timeform{0D}$).
During the observations, $T_\mathrm{sys}$ was 250--600 K  for  $^{12}$CO and 180--350 K for $^{13}$CO and C$^{18}$O,  respectively.
The average rms noise levels in $T_{\mathrm{mb}}$ scale with a velocity resolution of 1.3 km~s$^{-1}$ were $\sim$1.47 K for $^{12}$CO  and $\sim$ 0.69 K for $^{13}$CO and C$^{18}$O 
 with $\timeform{8''.5}$ pixel in the 1st quadrant regions (A and B), and $\sim$1.10 K for $^{12}$CO and $\sim$ 0.56 K for $^{13}$CO and C$^{18}$O with $\timeform{15''}$ pixel
  in the 3rd quadrant regions (C1, C2, and C3).

In the following subsection, we present the overall  distribution and kinematics of the region A in the 1st quadrant region and the individual regions of M17 and W51.

\begin{figure*}[htbp]

\includegraphics [width=16cm]{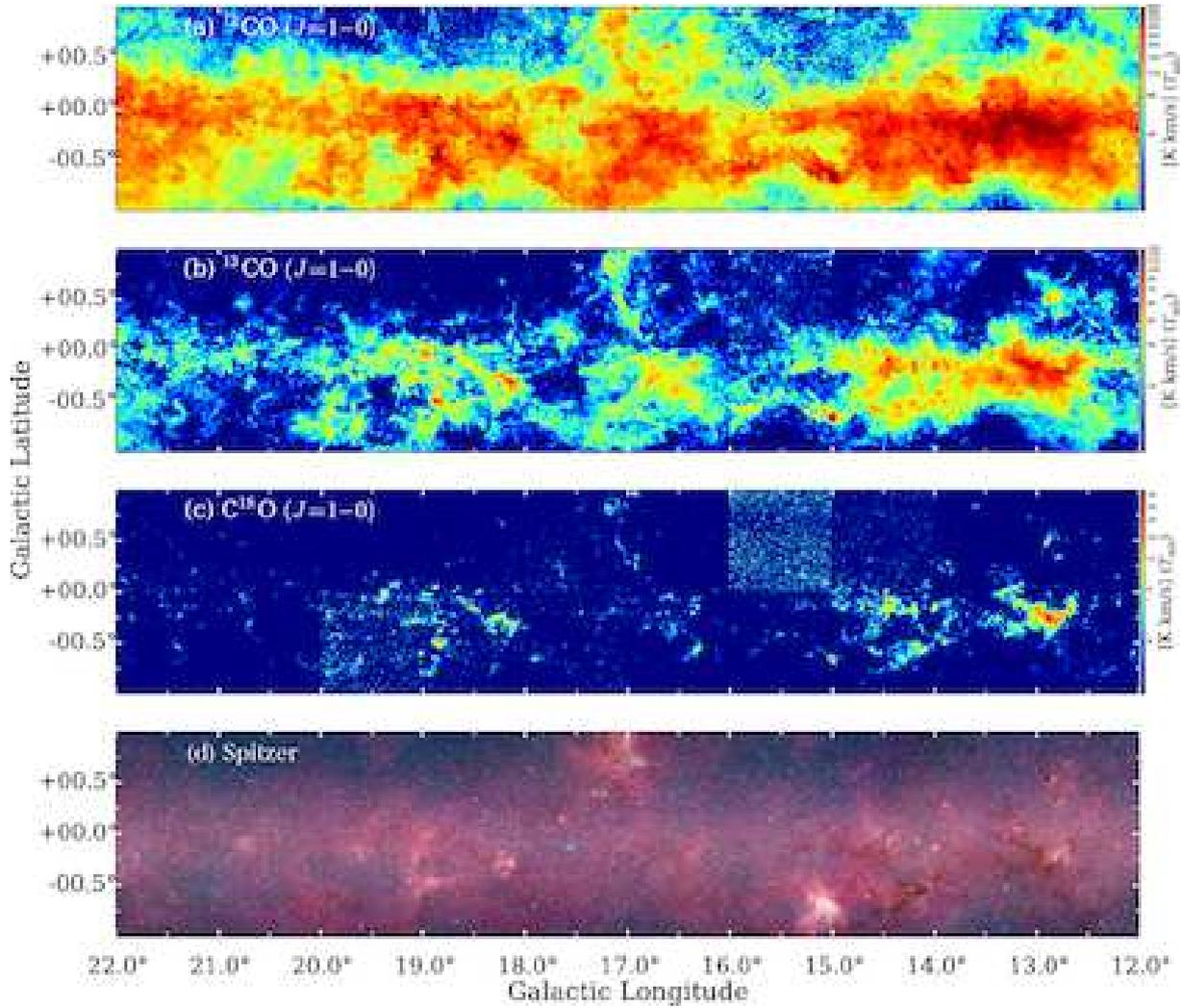}\\

\caption {Integrated  intensity maps of (a)$^{12}$CO, (b)$^{13}$CO and (c)C$^{18}$O $J=1-0$  in the region A ($l= \timeform{12D}$ to $\timeform{22D}$, $b =  \timeform{-1D}$ to $\timeform{1D}$) Integrated velocity range is  $-50$ km~s$^{-1}$ $<$ $V_\mathrm{LSR}$ $<$ 200 km~s$^{-1}$.  Lowest panel (d) shows the $Spitzer$ GLIMPSE image (blue:  3.6 $\mu$m, green: 5.4 $\mu$m, red: 8.0 $\mu$m) in the same region.}	
\label{fig3}
\end{figure*}

\subsection{Overall distribution and kinematics of molecular gas}

A high spatial dynamic range of our survey allows us to see details of large-scale structures of molecular clouds, e.g., filaments that have not been seen in previous surveys.
In figure 3,  the integrated intensity maps of $^{12}$CO, $^{13}$CO, and C$^{18}$O in the region A  are shown.
In this region, we find the giant molecular clouds associated with famous star-forming regions of M17 ({\it l}  $\sim$$\timeform{15.D1}$,  {\it b} $\sim$$\timeform{-0.D7}$) at a distance of 2.04~kpc \citep{chibueze16} (see section 4.2.1) 
and M16 ({\it l} $\sim$$\timeform{17.D0}$, {\it b} $\sim$$\timeform{+0.D8}$) at a distance of 1.8 kpc in the Sagittarius spiral arm \citep{bonatto06}.
Furthermore, we also find  isolated small molecular clouds with dense clumps detected in C$^{18}$O, e.g., located at {\it l}  $\sim$$\timeform{18D.25}$,  {\it b} $\sim$$\timeform{+0.D7}$
thanks to the non-biased mapping survey with both wide range and high angular resolution.
We reveal the distribution of molecular gas from diffuse and low-density regions traced by $^{12}$CO emission to compact and dense regions traced by $^{13}$CO and C$^{18}$O emission at once. 
We found that molecular gas, which is bright in $^{12}$CO, coincides well with bright nebulous regions seen in the $Spitzer$ GLIMPSE image. 
Indeed, well-defined shells are seen in both the GLIMPSE image and the $^{12}$CO/$^{13}$CO (e.g., the shell located at  {\it l} $\sim$$\timeform{15.D7}$,  {\it b} $\sim$$\timeform{-0.D5}$, which is more clearly seen in the velocity range of 50--60 km~s$^{-1}$ in figure 5 and figure 6). 
On the other hand, infrared dark clouds/filaments seen in the GLIMPSE image correspond well with the dense molecular clouds seen especially in the C$^{18}$O map.

A composite three-color image of peak intensity of $^{12}$CO, $^{13}$CO, and C$^{18}$O emission 
toward the region A is shown in figure 4.
The peak intensities of  $^{12}$CO, $^{13}$CO, and C$^{18}$O  are  presented in red, green and blue, respectively.
This three-color map represents the intensity ratios of the three CO lines, and it is possible to investigate 
the global physical conditions of Galactic GMCs in a way similar to \citet{barnes15}.
In figure 4, dense and warm regions are shown in white, where all $^{12}$CO, $^{13}$CO, and C$^{18}$O are intense. 
The red clouds, which are very bright in $^{12}$CO but very weak in $^{13}$CO and C$^{18}$O, indicate low optical depth 
and high temperature regions. On the other hand, green and blue clouds, where  $^{13}$CO/  $^{12}$CO or C$^{18}$O/ $^{12}$CO is high, are the high optical depth and low temperature, namely high-density and cold regions.

\begin{figure*}[htbp]
\begin{center}
\includegraphics [width=16cm]{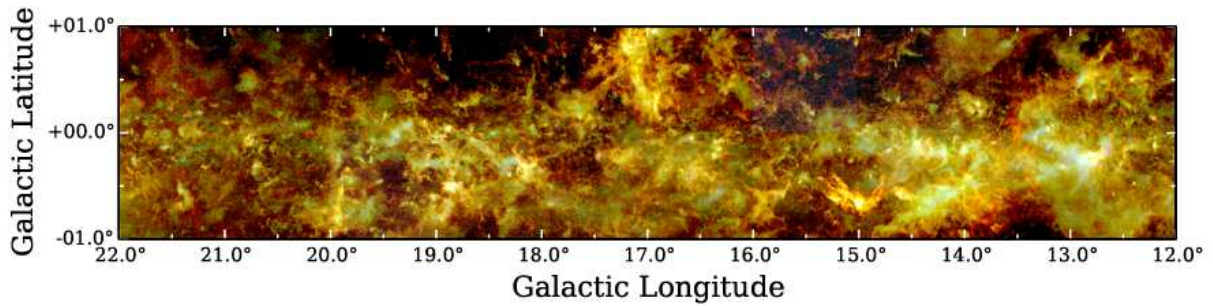}

\caption {Three-color  peak  $T_{\mathrm{mb}}$ intensity image of  the region A: $^{12}$CO (red), $^{13}$CO (green) and C$^{18}$O (blue).}
\label{fig4}
\end{center}
\end{figure*}

Figures 5--7 show channel maps of  $^{12}$CO,  $^{13}$CO, and C$^{18}$O emission with velocity intervals of 10 km~s$^{-1}$ in the region A.
We note that strips on the C$^{18}$O channel maps from 130 to 160 km~s$^{-1}$ in figure 7 are due to  artificial signals that are not fully removed at this moment.
However, these will be improved in the final products that will be released to the public.
The channel maps can separate the features along the line of sight into distinct clouds, which are not seen as prominent components in the integrated intensity maps.
We find extended diffuse emission features with the size of a few square degrees in the lower velocity maps (e.g., $V_\mathrm{LSR}$ = 0 to 10 km~s$^{-1}$)  in figure 5, 
whereas small, compact emission features in the higher velocity maps (e.g.,  $V_\mathrm{LSR}$ = 40 to 50 km~s$^{-1}$) in figure 5. 
This difference is mainly due to the difference of the distance of the clouds,
namely the latter is likely to be more distant than the relatively local emission in the 0-10 km~s$^{-1}$ interval, which can be estimated using kinematic distance derived from the Galactic rotation curve \citep{honma12, reid14}.

We find that most molecular clouds at  $l = \timeform{12D}$ to $\timeform{22D}$ are located at negative latitudes ($b < 0 \timeform{D}$) in the low-velocity range, while molecular gas in the high-velocity range is located at $b = 0 \timeform{D}$ (e.g., $V_\mathrm{LSR}$ = 100 to 130 km~s$^{-1}$  in figure 5).
This is due to the distance effect as follows.
Because the Sun lies 25~pc above the physical Galactic midplane \citep{juric08, bland16}, the true Galactic plane is likely to appear at negative latitudes.
On the other hand, the true Galactic plane will converge to $b = 0 \timeform{D}$ at a large distance.
The objects on the true Galactic plane at a distance of 2 kpc will be located at {\it b} = $\timeform{-0.D72}$, which is consistent with the location of M17SW at a distance of 2.04 kpc \citep{chibueze16}.
Therefore, M17SW is thought to be truly located on the physical Galactic plane.
We note that M16 ({\it l} $\sim$$\timeform{17D}$, {\it b} $\sim$$\timeform{+0.D5}$) departs from the tendency of the distribution of molecular clouds.
Thus M16 at a distance of 1.8~kpc is  located at $\sim$41~pc above on the physical Galactic plane, which is about 1/3 of the thickness of the molecular gas layer at 8.5 kpc from the Galactic Center (\cite{sanders84}; see \cite {heyer15}, and references therein).

\begin{figure*}[htbp]
\begin{center}
\includegraphics [width=14cm]{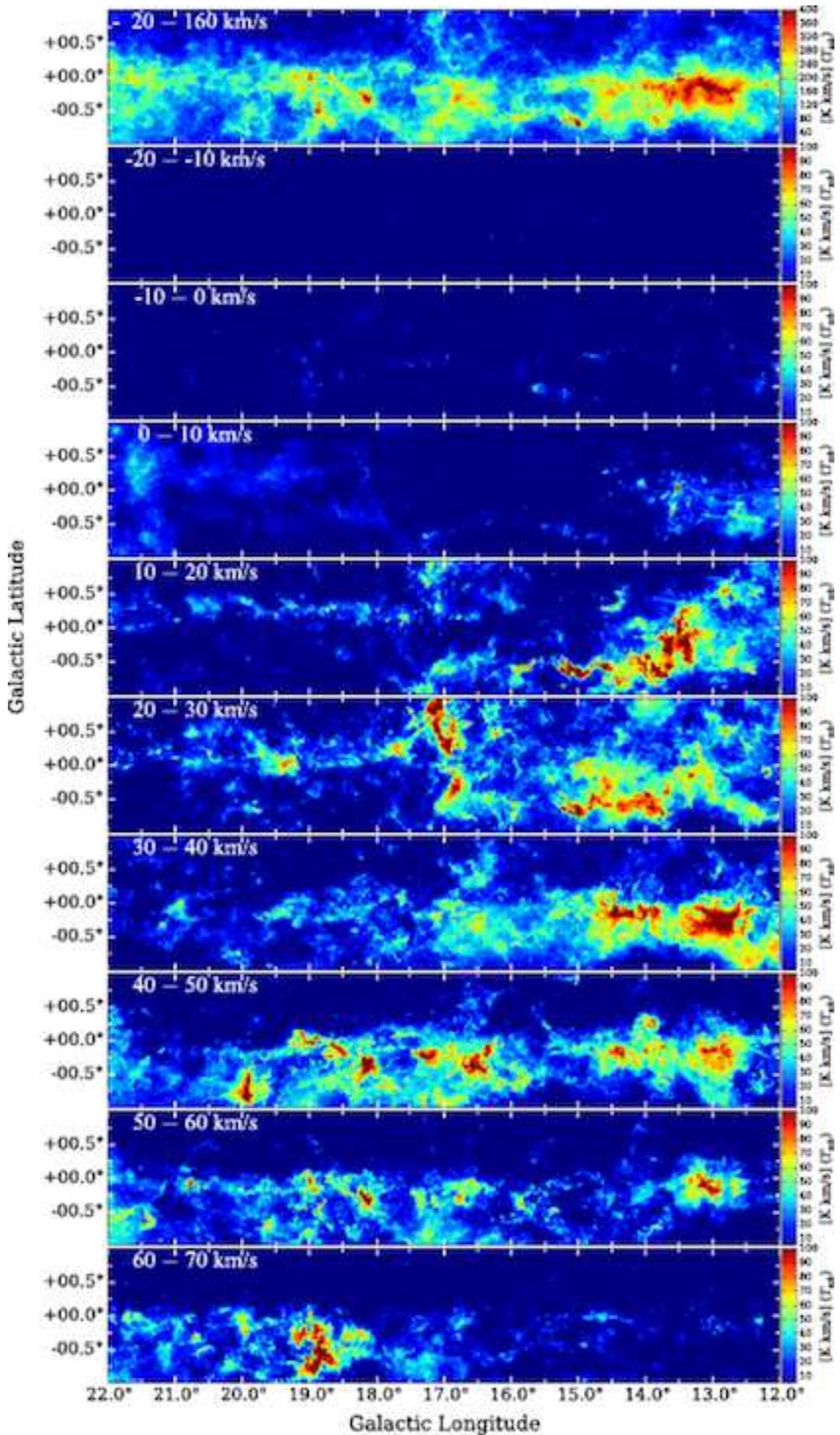}

\caption {Channel maps of $^{12}$CO in the region A with velocity intervals of 10 km~s$^{-1}$. The integrated intensity map is also shown at the top left.}
\label{fig5}
\end{center}
\end{figure*}

\addtocounter{figure}{-1}

\begin{figure*}[htbp]
\begin{center}
\includegraphics [width=14cm]{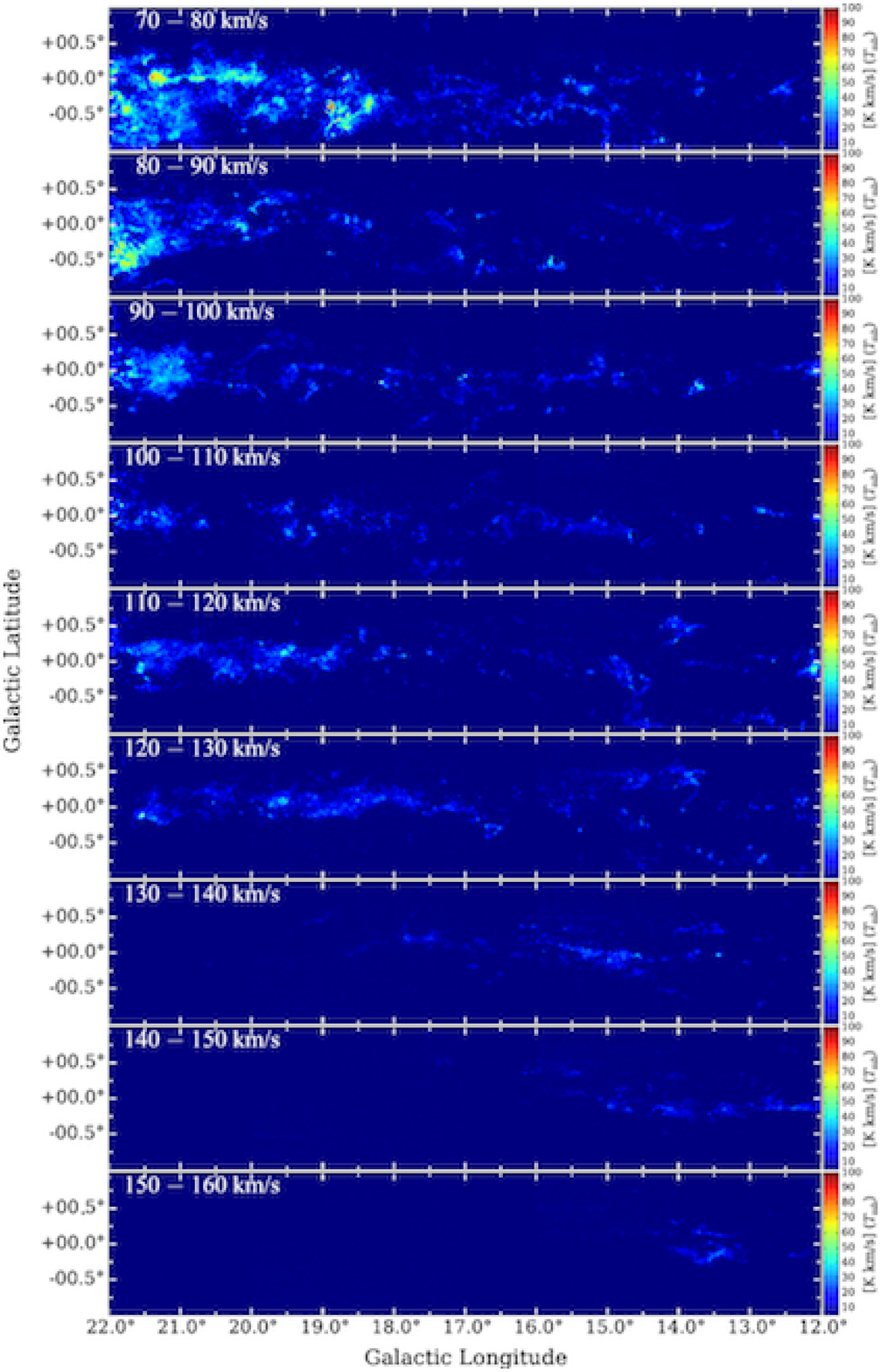}

\caption {(Continued)}
\label{fig5}
\end{center}
\end{figure*}

\begin{figure*}[htbp]
\begin{center}
\includegraphics [width=14cm]{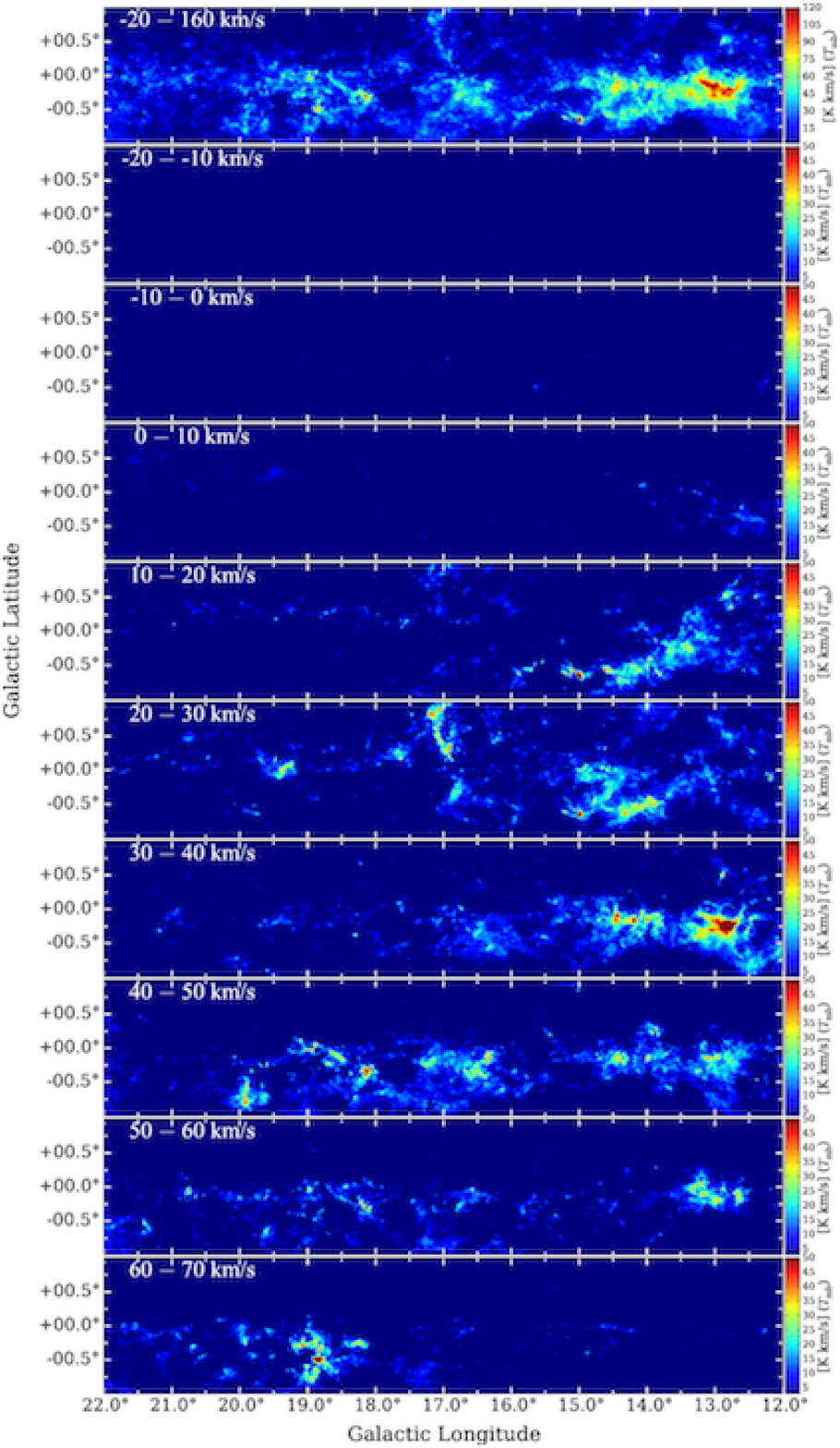}

\caption {Channel maps of $^{13}$CO in the region A with velocity intervals of 10 km~s$^{-1}$. The integrated intensity map is also shown at the top left.}
\label{fig6}
\end{center}
\end{figure*}

\begin{figure*}[htbp]
\begin{center}
\includegraphics [width=14cm]{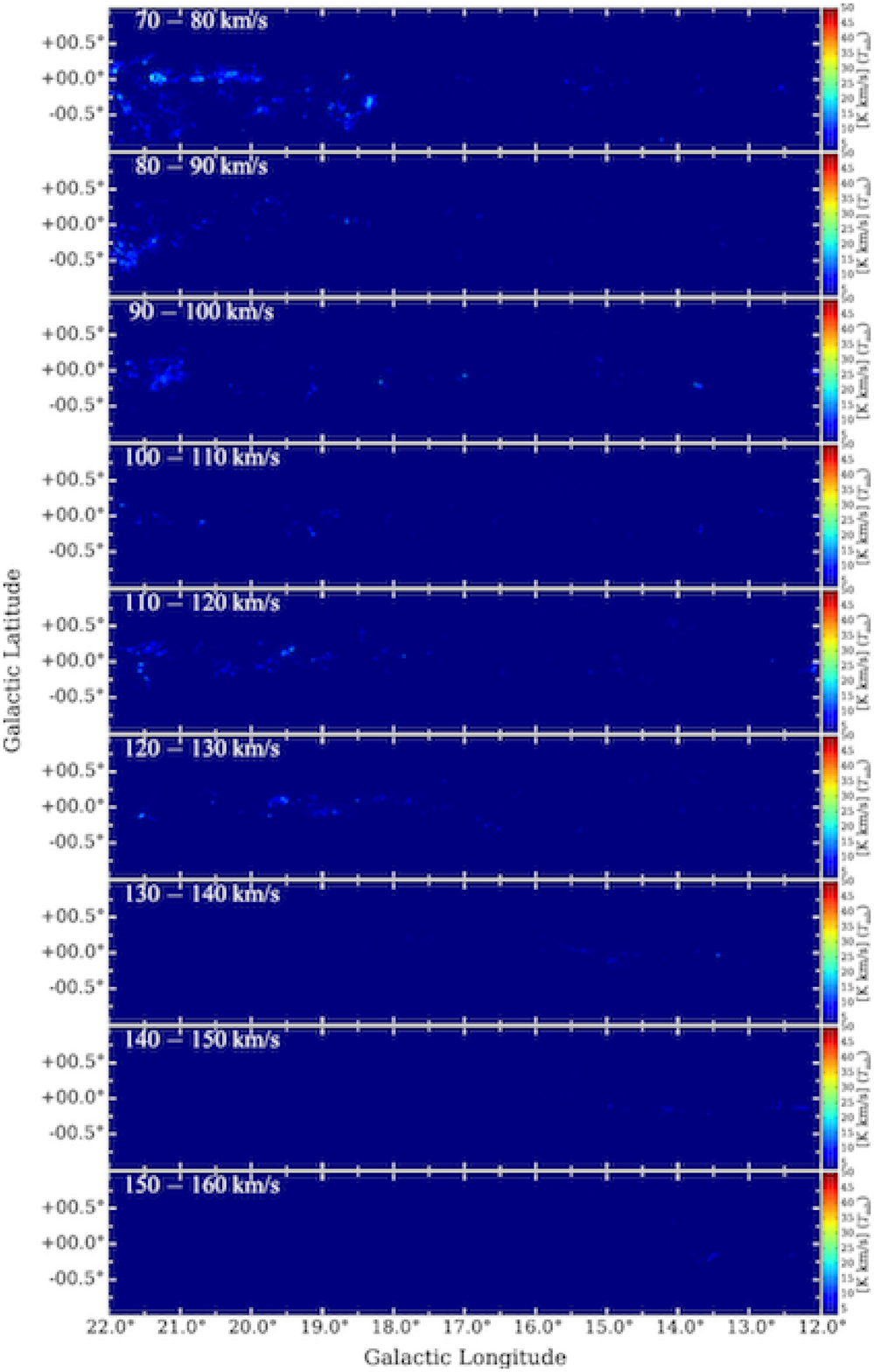}

\addtocounter{figure}{-1}

\caption {(Continued)}
\label{fig6}
\end{center}
\end{figure*}

\begin{figure*}[htbp]
\begin{center}
\includegraphics [width=14cm]{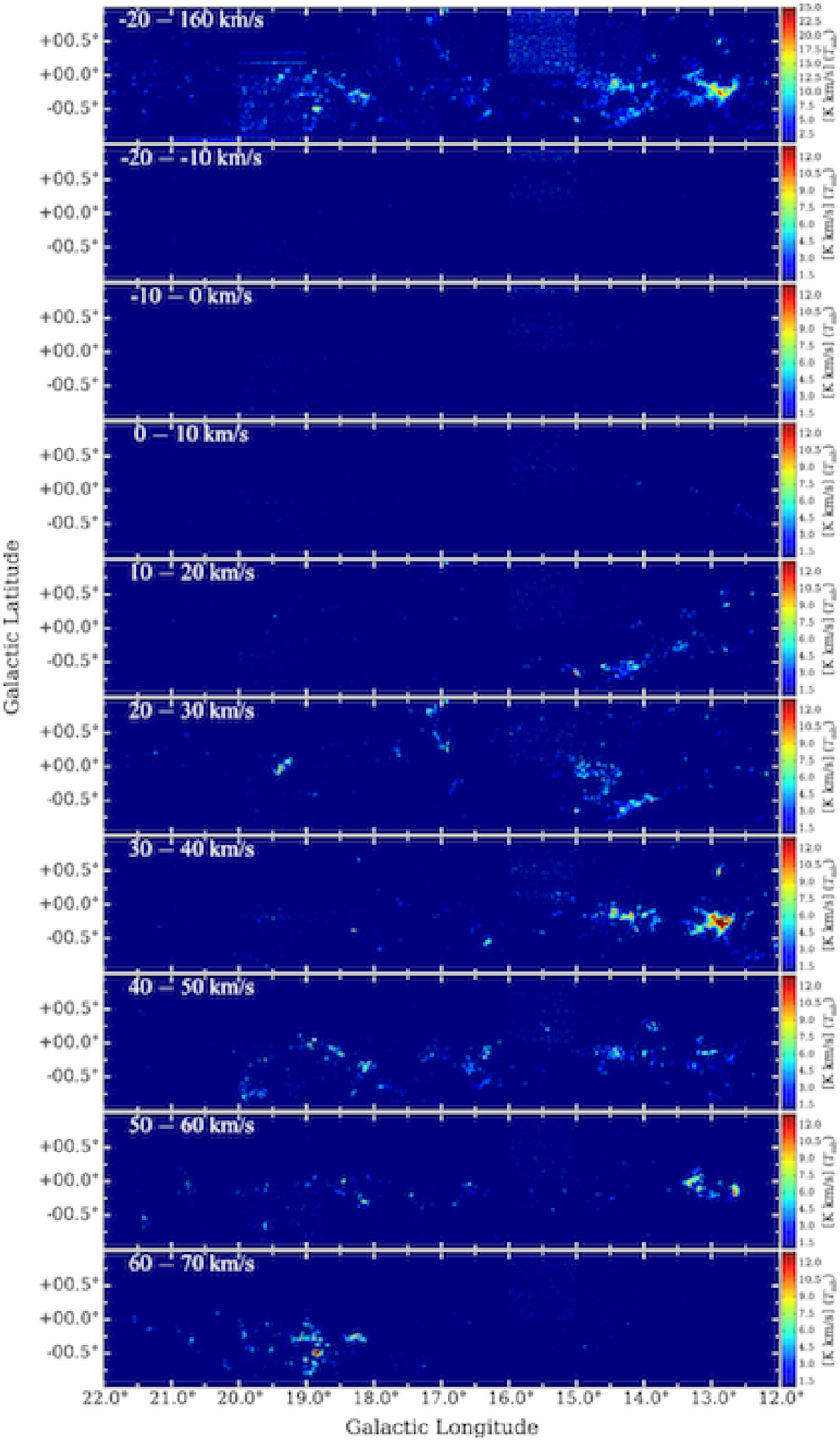}

\caption {Channel maps of C$^{18}$O in the region A with velocity intervals of 10 km~s$^{-1}$. The integrated intensity map is also shown at the top left.}
\label{fig7}
\end{center}
\end{figure*}

\begin{figure*}[htbp]
\begin{center}
\includegraphics [width=14cm]{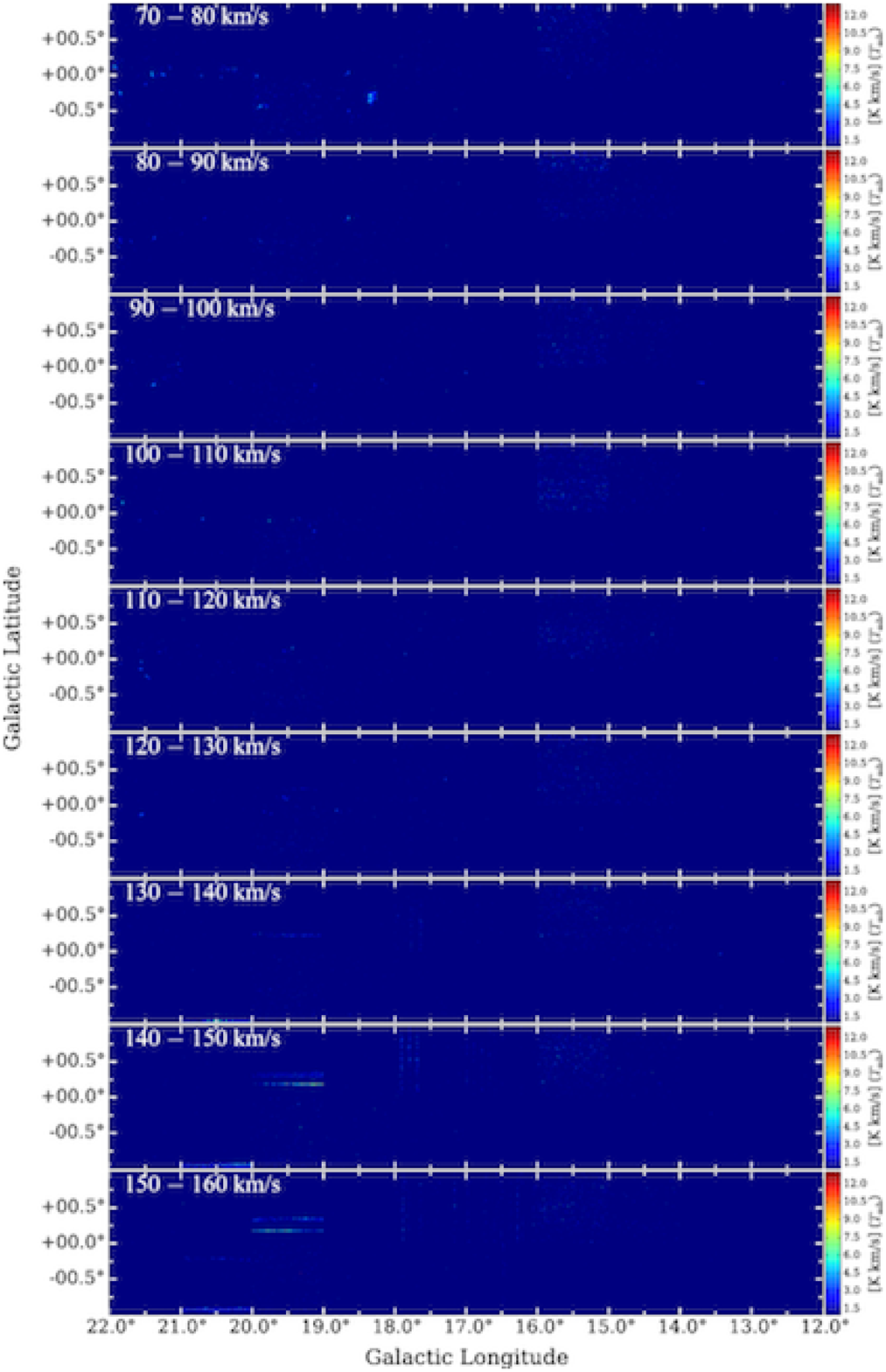}

\addtocounter{figure}{-1}

\caption {(Continued)}
\label{fig7}
\end{center}
\end{figure*}

Comparing the GLIMPSE image and GRS $^{13}$CO data \citep{jackson06}, \citet{ragan14} identified seven giant molecular filaments (GMFs) as both spatial and velocity coherent structure in the Galactic plane.
We can identify two GMFs within the region A.   One of the GMFs, GMF~18.0-16.8  which is located at $(l, b)$ = ($\timeform{16.D8}$, $\timeform{+0.D1}$) to ($\timeform{17.D3}$, $\timeform{+0.D6}$) and ($\timeform{17.D3}$, $\timeform{+0.D6}$) to ($\timeform{18.D0}$, $\timeform{0.D0}$) in the velocity range of 20--30 km~s$^{-1}$, is obvious in figure 5. Another one, GMF~20.0-17.9 which is located at $(l, b)$ = ($\timeform{17.D9}$, $\timeform{-0.D6}$) to ($\timeform{19.D2}$, $\timeform{0.D0}$)  in the velocity range of 40--50 km~s$^{-1}$, is prominent in figure 6.
\citet{tackenberg13} already identified the part of GMF~20.0-17.9 as the G18.93-0.03 infrared dark filament.
Because of non-biased mapping survey over a wide range, we can definitely identify  not only $^{13}$CO molecular filaments  such as GMFs corresponding to the IR dark filaments but also many molecular filaments without corresponding IR dark filaments.
For example, we can find the molecular filaments located at $l= \timeform{19D.8}$ to $\timeform{21D.5}$,  $b\sim\timeform{0D}$ in the velocity range of 70--80 km~s$^{-1}$ 
and $l= \timeform{14D.2}$ to $\timeform{15D.0}$,  $b\sim\timeform{0D}$ in the velocity range of 60--70 km~s$^{-1}$ in figure 5 and figure 6.
We can detect molecular emission from clouds/filaments even at a large distance, while it is difficult to identify distant IR dark clouds/filaments due to the strong foreground IR emission.
We detected C$^{18}$O emission as a high-density tracer toward GMF~20.0-17.9 at a distance of 3.6 kpc \citep{tackenberg13}.
So it seems that only very high-density regions of nearby $^{13}$CO filaments  are seen in the GLIMPSE image as IR dark filaments.

Figure 8 shows the longitude-velocity $(l-v)$ diagrams of the  $^{12}$CO,  $^{13}$CO, and C$^{18}$O emission in the region A.
These diagrams were made by integrating the emission over the full Galactic latitude range. 
Thanks to our high angular resolution, we find some interesting structures. 
For example, we find a remarkable structure connecting spiral arms from ($\timeform{20.D5}$, 20 km~s$^{-1}$) to ($\timeform{21D.5}$, 40 km~s$^{-1}$)  in the $l-v$ diagram of $^{12}$CO.
We find the narrow line emission feature of  $V_\mathrm{LSR}$ = 4 to 6 km~s$^{-1}$ located at  {\it l} = 17$^\circ$ to 22$^\circ$.
This feature must be real because it can be seen also in both $l-v$ diagrams of FUGIN $^{13}$CO and GRS, 
which corresponds to a part of the Aquila Rift  \citep{dame01, reid16} that is the local diffuse extended gas as shown in the previous channel map of $V_\mathrm{LSR}$ = 0 to 10 km~s$^{-1}$  in figure 5.
It is also interesting that the velocity width of the local diffuse gas increases from $l= \timeform{17D}$ to $\timeform{22D}$.
Indeed, we can find weak $^{12}$CO emission at a velocity range of 115--155  km~s$^{-1}$ located at  {\it l} = 13$^\circ$ to 22$^\circ$, 
which corresponds to the distant clouds located at $b\sim0\timeform{D}$as seen in the channel map of $V_\mathrm{LSR}$ = 110 to 160 km~s$^{-1}$  in figure 5.
We note that the material between 115--155 km~s$^{-1}$ and  {\it l} = 13$^\circ$ to 22$^\circ$ is located at the tangent point.

The $l-v$ diagrams reveal large-scale coherent velocity features that correspond to individual molecular cloud complexes.
The spiral arms, such as the Sagittarius, Scutum, and Norma arms,  can be seen at the positive velocity range in the $l-v$ diagrams 
 as estimated by \citet{ reid14}, respectively.
These emission features are much clearer in FUGIN $^{13}$CO and also C$^{18}$O $l-v$ diagrams than in \citet{dame01}.
Strong emission peaks are well aligned with the Sagittarius arm in this Galactic longitude range.
Although some peaks are departed from the Scutum  arm at {\it l} = $\timeform{18.2D}$, 
strong emission peaks are well aligned with the Scutum  arm within $\pm$5 km~s$^{-1}$.
Besides the main spiral arms, we can see there is a structure that crosses the Sagittarius arm at  $(l, v)$ = ($\timeform{14D}$, 22 km~s$^{-1}$)
with a velocity gradient from 10 to 30 km s$^{-1}$ at $l = \timeform{12D.5}$  to $\timeform{15D.5}$ in the $l-v$ diagram of $^{13}$CO.
We also find the aligned emission at  $(l, v)$ = ($\timeform{12.5D}$, 55 km~s$^{-1}$) to  ($\timeform{18.5D}$, 70 km~s$^{-1}$).
We note that there is weak $^{12}$CO emission at a negative velocity of $\sim$$-4$ km~s$^{-1}$, which may belong to the far side of the Perseus arm \citep{dame01, reid16}.

\begin{figure}[htbp]
\begin{center}
\includegraphics [width=8.2cm]{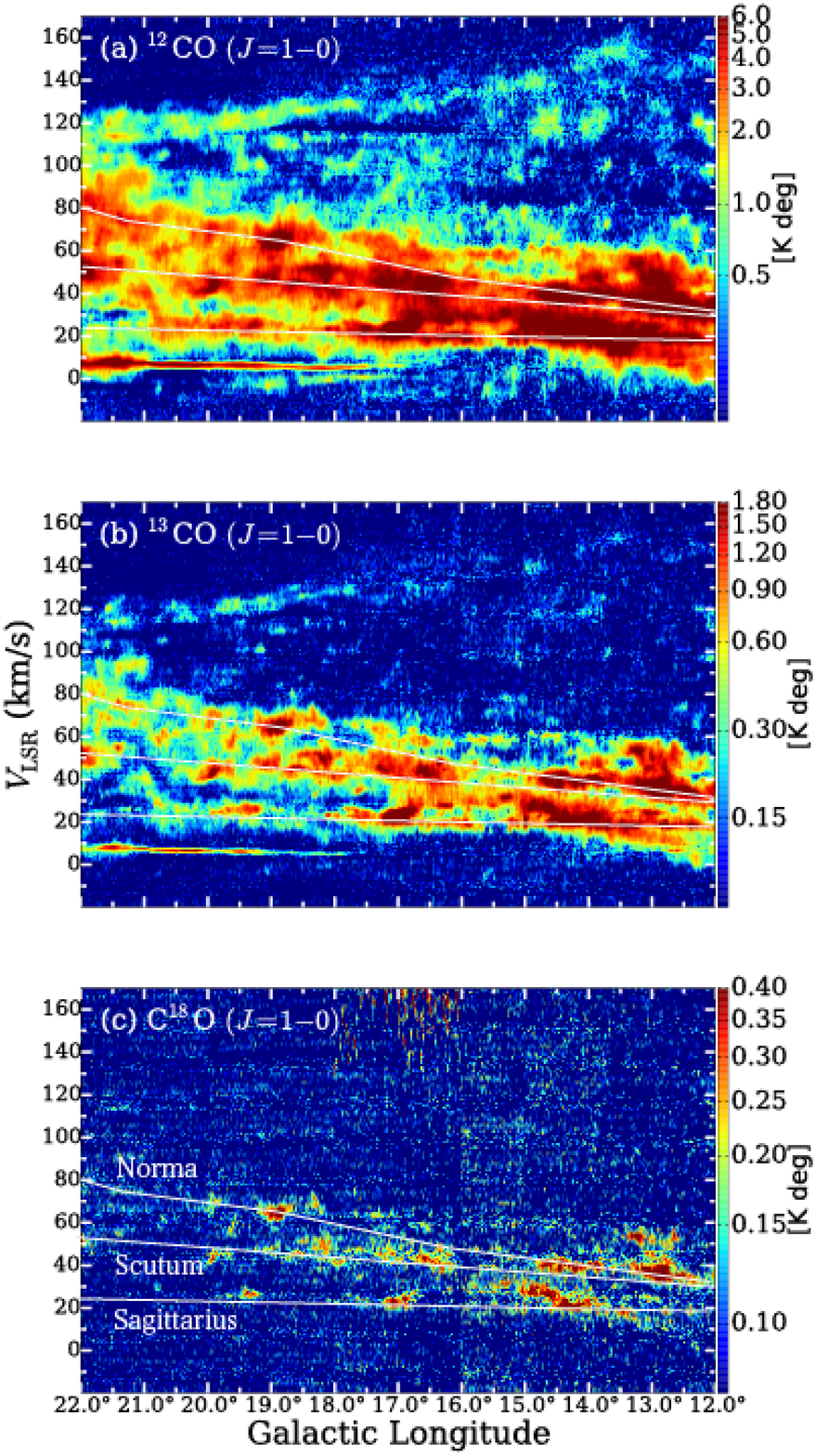}
\end{center}
\caption {Longitude-velocity diagrams of the $^{12}$CO,  $^{13}$CO, and C$^{18}$O emission in the region A. These diagrams have been made by collapsing over the  latitude range for each longitude.  The solid lines indicate the spiral arm loci of  Norma, Scutum, and Sagittarius as estimated by \citet{ reid14}.}
\label{fig8}
\end{figure}

\subsection{Individual Sources}

In this subsection, we present a close-up view of interesting sources included in our mapping area, M17 and W51.

\begin{figure*}[htbp]
\begin{center}
\includegraphics [width=12cm]{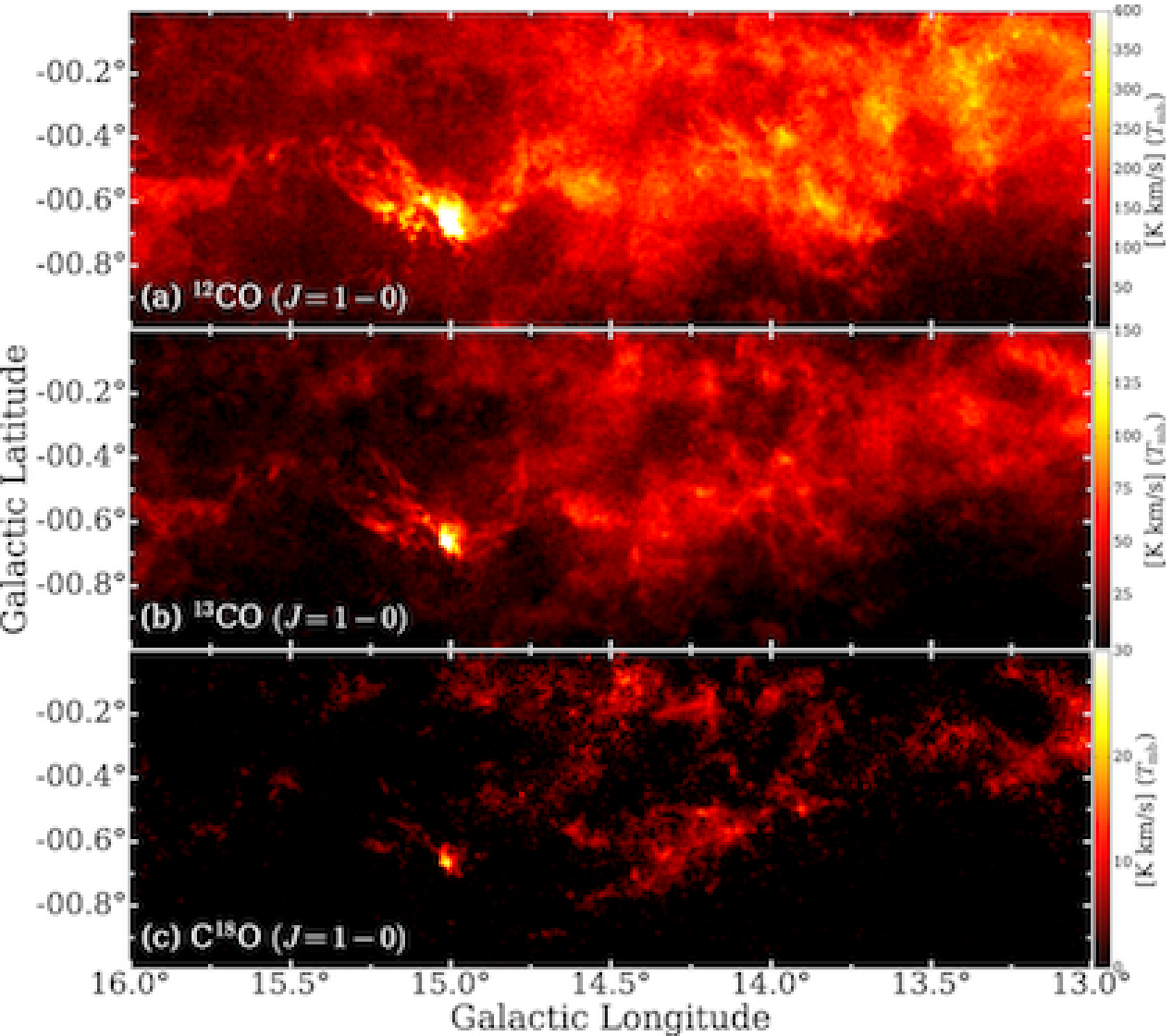}
\end{center}
\caption {Integrated intensity maps of M17 in $^{12}$CO, $^{13}$CO, and C$^{18}$O.
The integrated velocity range is $10$~km s$^{-1}$ -- 40~km s$^{-1}$.}
\label{fig9}
\end{figure*}

\begin{figure*}[htbp]
\begin{center}
\includegraphics [width=12cm]{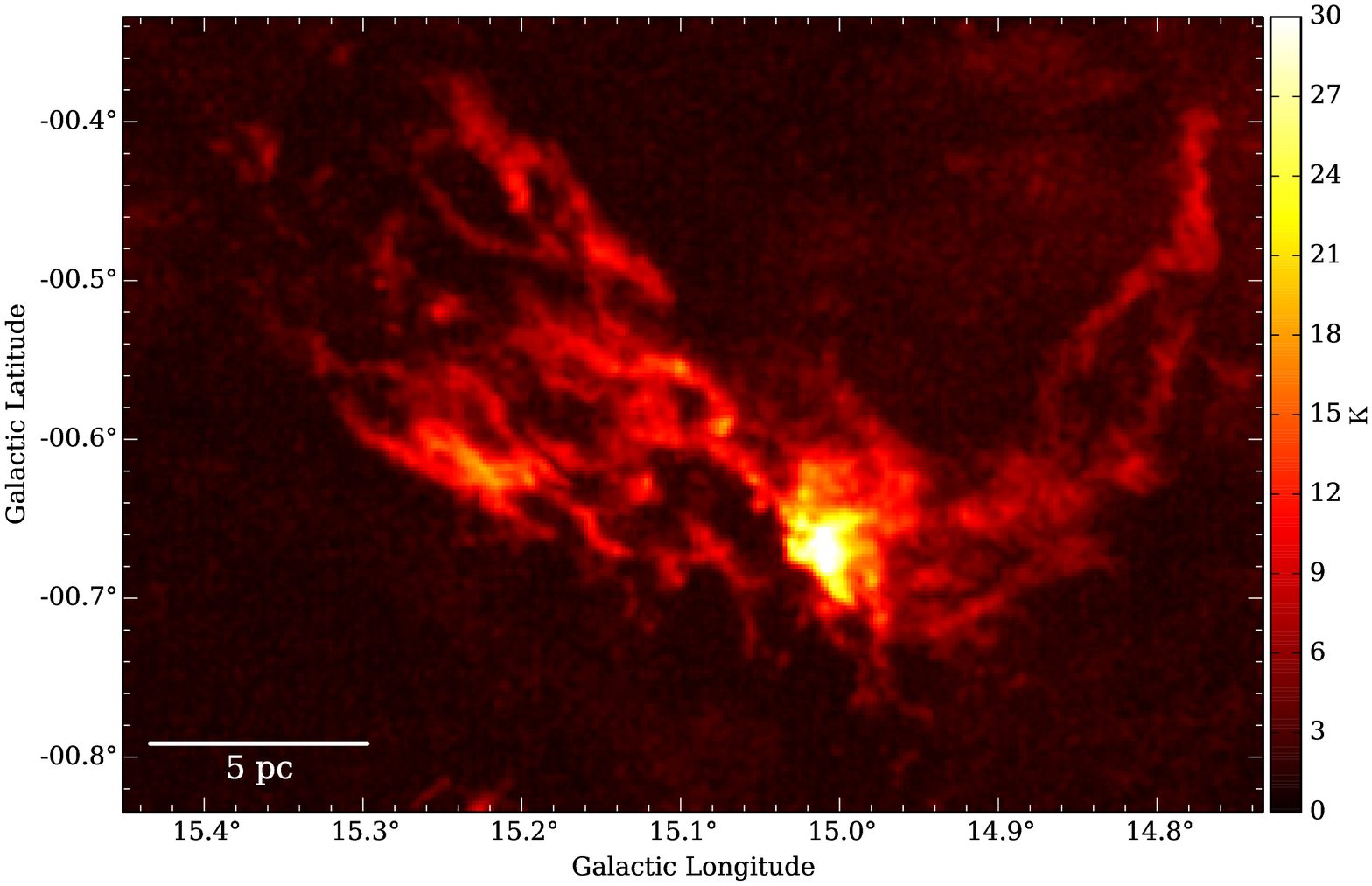}

\caption {Close-up view of M17SW molecular cloud in $^{13}$CO peak  $T_{\mathrm{mb}}$ intensity within a velocity range of 10~km s$^{-1}$ -- 40~km s$^{-1}$.}
\label{fig10}
\end{center}
\end{figure*}

\subsubsection{M17}

M17 is a well-known H{\sc ii} region at a distance of 1.98~kpc  \citep{xu11} and 2.04 kpc  from the new VERA  parallax distance measurement \citep{chibueze16} lying on the Carina-Sagittarius arm. 
The  spatial resolution of FUGIN, 20$^{\prime\prime}$,  corresponds to $\sim 0.1$ pc at the distance of 2.04~kpc, which can resolve  individual high-density cores.
The associated molecular cloud mass is estimated to be $2 \times 10^5 M_\odot$ with a size of 70 pc $\times$ 15 pc \citep{elmegreen79}.
This region contains M17SW molecular cloud ($l\sim\timeform{15D.0}$, $b\sim\timeform{-0.D65}$), where intense star-formation occurs, and M17SWex 
($l\sim\timeform{14D.3}$, $b\sim\timeform{-0.D6}$), which is separated by $\sim$50 pc to the southwest of M17. 

From $Spitzer$/IRS mid-infrared spectral maps of the M17SW region as well as IRSF/SIRIUS Br$_{\gamma}$ and FUGIN  $^{13}$CO data, 
 \citet{yamagishi16} found that the PAH  (polycyclic aromatic hydrocarbon) emission features are bright in the region between the HII region clearly traced by Br$\gamma$ and the molecular cloud traced
by $^{13}$CO due to both the high spatial resolution of the $^{13}$CO map and the small extinction in the Br$\gamma$ map, supporting that the PAH emission  mostly originates in photo-dissociation regions.

In contrast to M17SW, the latter hosts a smaller number of massive stars, implying that M17SWex is in an earlier phase than M17SW \citep{povich10, povich16}.
M17SWex is also reported as the IRDC, G15.225-0.506, consisting of ``hub filaments'' system  \citep{busquet13, ohashi16}. 
On the east side of M17SW ($l\sim\timeform{15D.7}$,  $b\sim\timeform{-0.D6}$), there is M17EB exhibiting an advanced  star-formation \citep{povich09}. 
Thus, M17 hosts multiple star-forming regions at various evolutionary stages. 

Figure 9 displays integrated intensity maps of $^{12}$CO, $^{13}$CO, and C$^{18}$O toward $3^\circ \times 1^\circ$ region ($ l=13^\circ$ to 16$^\circ$,  $b = -1^\circ$ to 0$^\circ$)  containing M17SW, SWex and EB. 
We find that $^{12}$CO and $^{13}$CO emission lines are widespread across the observed region, whereas C$^{18}$O, which traces dense molecular gas, is rather confined to clumps with multiple peaks. 
This indicates that high-density molecular gas condensations exist within the extended molecular clouds. 

Very intense $^{12}$CO emission are found as well as significant peaks of $^{13}$CO and C$^{18}$O in the M17SW region with the vigorous star formation.
We note that multiple filaments seen in $^{12}$CO and $^{13}$CO snuggle up to each other. A close-up view of the entangled filaments is  shown in figure 10. 
A high spatial dynamic range of our survey with three times higher angular resolution has revealed  filaments that have not been seen in previous surveys.
On the other hand, no significant peak in $^{12}$CO is seen in the M17SWex region, although we identify several significant peaks in C$^{18}$O. 
C$^{18}$O emission distribution is similar to the $Spitzer$ dust image in the M17SWex region.

We see little C$^{18}$O emission in M17EB, which is claimed to be in an advanced stage of star-formation, although significant $^{12}$CO and $^{13}$CO emission lines are observed. 
These results suggest that distributions of each line differ from region to region, presumably reflecting their evolutionary stages of star formation. 
We note that wide-spread $^{12}$CO, $^{13}$CO, and C$^{18}$O emission lines near $b \sim 0^\circ$ regions have systematically different radial velocities from those in M17SW and SWex, indicating that these are not physically associated with M17 molecular clouds.

\subsubsection{W51}

W51 is one of the most active star-forming regions in the Galaxy including many HII regions (W51A, W51B) and  a supernova remnant (W51C) located in the Sagittarius arm.
Many massive protostars also have been found in W51 \citep{kang09}.
W51 GMC is one of the most massive GMCs in the Galaxy ($M_{\mathrm{gas}} \sim 1.2 \times 10^{6} M_{\odot}$)  \citep{carpenter98} associated with W51A and W51B. 
W51 GMC is known to be composed of components with different velocities.
The high star-formation activity of W51 is thought to be caused by a collision between the components \citep{carpenter98, okumura01}. 
Figure11 shows an integrated intensity map of W51 in $^{12}$CO, $^{13}$CO, and C$^{18}$O within the velocity ranges of 45--55~km s$^{-1}$, 55--65~km s$^{-1}$, and 65--75~km s$^{-1}$. 
The maps cover the whole region of W51 including W51A, W51B, and W51C.
The filamentary structure from east to west in the 65--75 km~s$^{-1}$ map is called the high-velocity stream (HVS). The HVS is also clearly seen in C$^{18}$O. 
In addition to the main components such as the HVS and W51 main which is the strong peak seen in the 55--65 km~s$^{-1}$ map, we can see that there is diffuse surrounding molecular gas around W51 in the $^{12}$CO maps.

\begin{figure*}[htbp]
\begin{center}
\includegraphics [width=16cm]{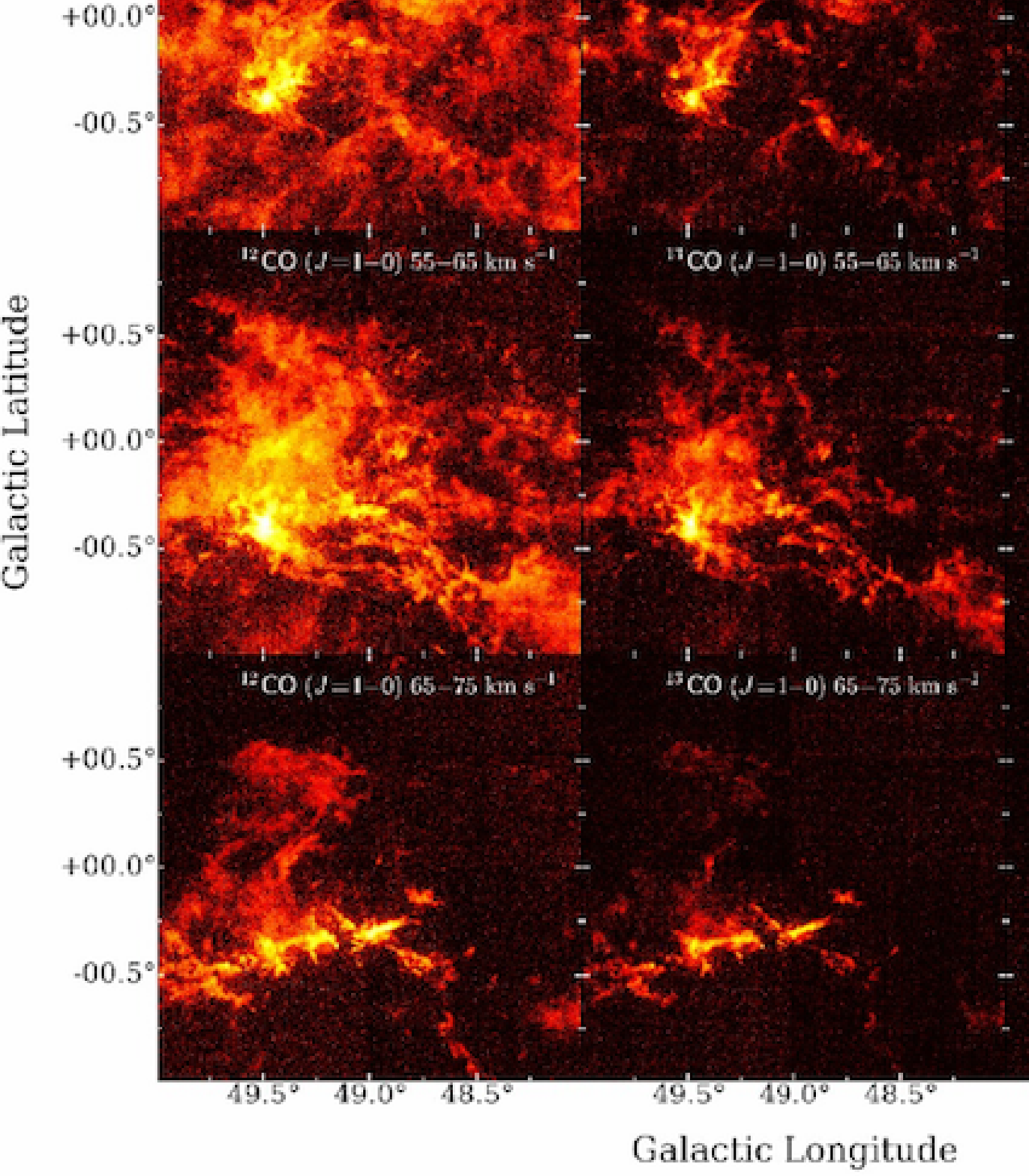}
\end{center}
\caption {Integrated intensity maps of W51 in $^{12}$CO (left), $^{13}$CO (middle), and C$^{18}$O ($J=1-0$) (right) lines.
The velocity ranges are 45--55~km s$^{-1}$(top), 55--65~km s$^{-1}$(middle), and 65--75~km~s$^{-1}$(bottom).}
\label{fig11}
\end{figure*}

Figure12 shows the intensity profiles from individual spectra of $^{12}$CO, $^{13}$CO, and C$^{18}$O toward the $^{12}$CO peaks at $(l, b)$=($\timeform{49.369D}, $\timeform{-0.353D}) and ($\timeform{49.494D}, $\timeform{-0.375D}) positions.
Even in the regions with complex profiles of $^{12}$CO and $^{13}$CO lines, C$^{18}$O line can be fitted with a single Gaussian, which is expected to correspond to a dense clump.
\citet{schloerb87} show a good general correlation between the 1.3 mm continuum, which traces the column density of dust, and the C$^{18}$O ($J=2-1$) line emission in the Orion, W49, and W51 molecular clouds.
\citet{rigby16} also mentioned that the integrated intensity maps of C$^{18}$O  ($J=3-2$) and ATLASGAL 870 $\mu$m image show a high degree of similarity in their spatial distributions.
These will support that C$^{18}$O ($J=1-0$)  is a good tracer of dense clumps.
Furthermore, we can  derive their physical properties by using the three lines and JCMT CO ($J=3-2$)  data \citep{dempsey13, rigby16} which have almost the same angular resolution as our data.
We will do that and make comparisons with the properties of YSOs such as distribution and mass function in a forthcoming paper.

\begin{figure}[htbp]
\begin{center}
\includegraphics [width=8cm]{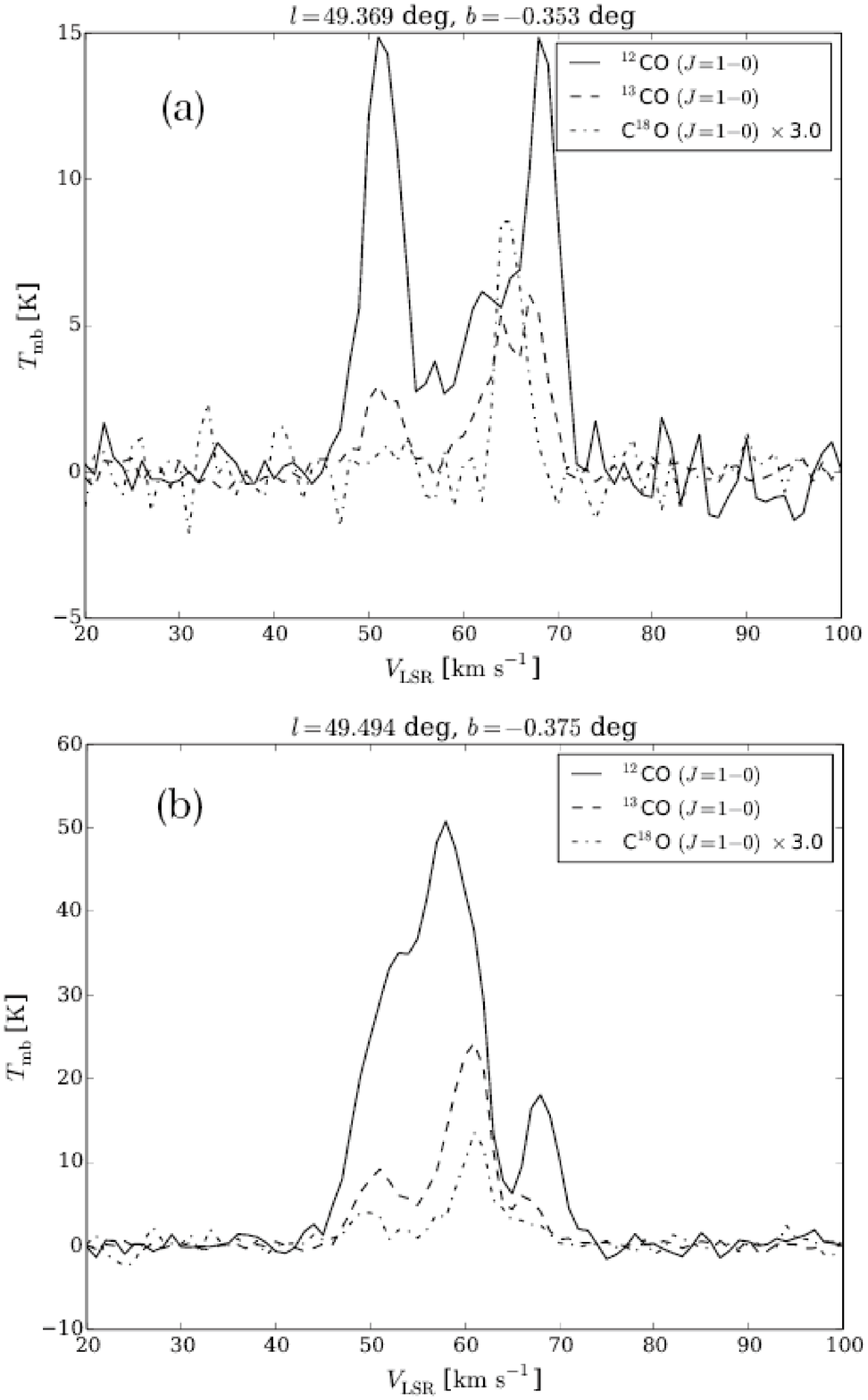}
\end{center}
\caption {Intensity profiles from individual spectra of $^{12}$CO(solid line),$^{13}$CO(dashed line) and C$^{18}$O(dotted line) at (a) $(l,b)$=($\timeform{49.369D}, $\timeform{-0.353D}) and (b) ($\timeform{49.494D}, $\timeform{-0.375D}). The vertical scale of C$^{18}$O is expanded by three times over others.}
\label{fig12}
\end{figure}

\section{Summary}

We present an overview of the FUGIN (FOREST Unbiased Galactic plane Imaging survey with the Nobeyama 45-m telescope) project, which is the Galactic plane survey with the Nobeyama 45-m telescope in three CO $J=1-0$ transitions ($^{12}$CO, $^{13}$CO, and C$^{18}$O).
These CO lines can be observed simultaneously with the new multi-beam FOREST receiver .
The mapping regions cover $10 \timeform{D} \leq l \leq 50 \timeform{D}$,  $|b| \leq \timeform{1D}$ and $198 \timeform{D} \leq  l  \leq 236 \timeform{D}$,  $|b| \leq \timeform{1D}$.

In the observing term from March 2014 to May 2015, we have covered the regions of the 1st quadrant $l= \timeform{12D}$ to $\timeform{22D}$, $b =  \timeform{-1D}$ to $\timeform{1D}$ and $l= \timeform{44D}$ to $\timeform{50D}$, $b =  \timeform{-1D}$ to $\timeform{1D}$, also of the 3rd quadrant $l= \timeform{198D}$ to $\timeform{202D}$, $b =  \timeform{-1D}$ to $\timeform{1D}$, $l= \timeform{213D}$ to $\timeform{218D}$, $b =  \timeform{0D}$ to $\timeform{1D}$
and $l= \timeform{221D}$ to $\timeform{227D}$, $b =  \timeform{-1D}$ to $\timeform{0D}$.
The velocity ranges and velocity resolution are $-50$ to 200 km~s$^{-1}$  for the 1st quadrant and $-50$ to 150 km~s$^{-1}$ for the 3rd quadrant, and 1.3 km~s$^{-1}$, respectively.

We also present the initial results of the FUGIN project in the region of the 1st quadrant $l= \timeform{12D}$ to $\timeform{22D}$, $b =  \timeform{-1D}$ to $\timeform{1D}$.
The data of $^{12}$CO, $^{13}$CO, and C$^{18}$O show a wide range and detailed structures of molecular clouds such as filaments that have not been seen in previous surveys thanks to the high angular resolution of the Nobeyama 45-m telescope.
The data demonstrate that we can get the distributions of diffuse and low-density gas traced by $^{12}$CO emission and compact high-density gas traced by $^{13}$CO and C$^{18}$O emission at the same time.
The ratios of the three CO lines indicate the variation of the physical conditions of molecular gas in various scales.
Three-dimensional spectral line data revealed  the  distinct clouds, which are not seen as prominent components in the integrated intensity maps, 
and large-scale kinematics of molecular gas like spiral arms, which are more clearly traced in $^{13}$CO and C$^{18}$O emission.

From a close-up view of  the M17 region,
we found that high-density molecular gas seen in C$^{18}$O is confined to multiple clumps within the extended molecular clouds traced by $^{12}$CO and $^{13}$CO. 
Multiple filaments like entangled strings are revealed, which have not been revealed in previous surveys in the M17SW region.
The FUGIN data revealed that the W51 molecular complex has the high-density clumps, including the filamentary structure seen in $^{13}$CO and C$^{18}$O is surrounded by diffuse molecular gas seen in $^{12}$CO.

As shown by the initial results, the final data obtained by the FUGIN project will help us to trace the evolution of the interstellar medium from extended low-density gas in the galactic large-scale to high-density clumps/cores in the small-scale which are the sites of star and cluster formation.

\begin{ack}
The authors would like to thank the members of the 45-m group of Nobeyama Radio Observatory for support during the observation.
Data analysis was carried out on the open use data analysis computer system at the Astronomy Data Center, ADC, of the National Astronomical Observatory of Japan.
\end{ack}







\begin{thebibliography}{}
\bibitem[Aguirre et al.(2011)]{aguirre11}
Aguirre, J. E., et al. 2011,
\apjs, 192, 4

\bibitem[Anderson et al.(2012)]{anderson12}
Anderson, L. D. et al. 2012,
\aap, 542, 10

\bibitem[Barnes et al.(2015)]{barnes15}
Barnes, P. J., Muller, E., Indermuehle, B., O'Dougherty, S. N., Lowe, V., Cunningham, M., Hernandez, A. K., \& Fuller, G. A. 2015,
\apj, 812, 6

\bibitem[Benjamin et al.(2003)]{benj03}
Benjamin, R. A.  et al. 2003,
\pasp, 115,  953

\bibitem[Bland-Hawthorn \& Gerhard (2016)]{bland16}
Bland-Hawthorn, J. \& Gerhard, O. 2016,
\araa, 54, 529 

\bibitem[Bonatto et al.(2006)]{bonatto06}
Bonatto, C.,  Santos, J. F. C., Jr., \& Bica, E. 2006,
\aap, 445, 567

\bibitem[Busquet et al.(2013)]{busquet13}
Busquet, G. et al. 2013,
\apjl, 764, L26

\bibitem[Carey et al.(2009)]{carey09}
Carey, S. J. et al.\ 2009,
\pasp, 121, 76

\bibitem[Carpenter \& Sanders(1998)]{carpenter98}
Carpenter, J. M., \& Sanders, D. B.\ 1998, 
\aj, 116, 1856

\bibitem[Chibueze et al.(2016)]{chibueze16}
 Chibueze, J.~O. et al.\ 2016, \mnras, 460, 1839 

\bibitem[Churchwell et al.(2006)]{churchwell06}
Churchwell, E., et al. 2006, \apj, 649, 759

\bibitem[Contreras et al.(2013)]{contreras13}
Contreras, Y. et al. 2013, 
\aap, 549, A45

\bibitem[Cyganowski et al.(2008)]{cyganowski08} 
Cyganowski, C. J. et al. 2008, 
\aj, 136, 2391

\bibitem[Dame, Hartmann and Thaddeus(2001)]{dame01}
Dame, T. M., Hartmann, D., \& Thaddeus, P. 2001, \apj, 547, 792

\bibitem[Dempsey et al.(2013)]{dempsey13}	
Dempsey, J. T., Thomas, H. S., \& Currie, M. J. 2013, 
\apjs, 209, 8

\bibitem[Doi et al.(2015)]{doi15}
Doi, Y. et al. 2015, 
\pasj, 67, 50

\bibitem[Elmegreen et al.(1979)]{elmegreen79}
Elmegreen, B. G., Lada, C. J., \& Dickinson, D. F. 1979, 
\apj, 230, 415

\bibitem[Goldreich \& Kwan(1974)]{goldreich74}
Goldreich, P., Kwan, J. 1974, 
\apj, 189, 441

\bibitem[Handa et al.(2012)]{handa12}
Handa, T., Yoda, T., Kohno, K.,  Morino, J.-i.,  Nakajima, T.,  Kuno, N.,  Ogawa, H., \&Kimura, K. 2012,
in \asp, 458,
Galactic Archaeology: Near-Field Cosmology and the Formation of the Milky Way, ed. W. Aoki, M. Ishigaki, T. Suda, T. Tsujimoto, \& N. Arimoto (San Francisco: ASP), 221

\bibitem[Heyer \& Dame(2015)]{heyer15}
Heyer, M. \& Dame, T.M. 2015,
\araa, 53, 583

\bibitem[Honma et al.(2012)]{honma12}
Honma, M. et al. 2012, \pasj, 64, 136

\bibitem[Ikeda \& Kitamura(2009)]{ikeda09}
Ikeda, N. \& Kitamura, Y. 2009, \apj, 705, L95

\bibitem[Jackson et al.(2006)]{jackson06}
Jackson, J. M. et al. 2006, \apjs, 163, 145

\bibitem[Juri\'c et al.(2008)]{juric08}
Juri\'c, M. et al. 2008. \apj,  673, 864
  
\bibitem[Kang et al.(2009)]{kang09}
Kang, M., Bieging, J. H., Povich, M. S. \&  Lee, Y. 2009, 
\apj, 706, 83

\bibitem[Kamazaki et al.(2012)]{kamazaki12}
Kamazaki, T. et al. 2012, 
\pasj, 64, 29

\bibitem[Kuno et al.(2011)]{kuno11}
Kuno, N. et al. 2011, 
General Assembly and Scientific Symposium,  XXXth URSI, \\
http://ieeexplore.ieee.org/xpl/articleDetails.jsp?arnumber=6051296

\bibitem[Mercer et al.(2005)]{mercer05}
Mercer, E. P. et al. 2005, 
\apj, 635, 560

\bibitem[Minamidani et al.(2016a)]{minamidani16a}
Minamidani, T. et al. 2016a, 
EAS Publications Series, 75-76, Conditions and Impact of Star Formation, ed. R. Simon, R. Schaaf, \& J. Stutzki (EDP Sciences), 193

\bibitem[Minamidani et al.(2016b)]{minamidani16b}
Minamidani et al. 2016b, 
Proc. SPIE 9914, 99141Z 
	
\bibitem[Molinari et al.(2010)]{molinari10}
Molinari, S. et al. 2010,
\aap, 518, 100

\bibitem[Moore et al.(2015)]{moore15}
Moore, T. J. T. et al. 2015, 
\mnras, 453, 4264

\bibitem[Nishimura et al.(2015)]{nis15} 
Nishimura, A. et al. 2015, \apjs, 216, 18 

\bibitem[Ohashi et al.(2016)]{ohashi16}
Ohashi, S., Sanhueza, P., Chen, H.-R. V., Zhang, Q., Busquet, G., Nakamura, F., Palau, A., \& Tatematsu, K. 2016, \apj, 833, 209

\bibitem[Okumura et al.(2001)]{okumura01}
Okumura, S.,  Miyawaki, R., Sorai, K., Yamashita, T., \& Hasegawa, T. 2001, 
\pasj, 53, 793

\bibitem[Onishi(2008)]{onishi08}
Onishi, T. 2008, 
Astrophysics and Space Science Proceedings, Mapping the Galaxy and Nearby Galaxies, ed. K. Wada, \&  F. Combes (Springer), 11

\bibitem[Onishi et al.(2013)]{onishi13}
Onishi, T. et al.  2013,
\pasj, 65, 78

\bibitem[Oyama et al.(2012)]{oyama12}
Oyama, T. et al. 2012, IVS 2012 General Meeting Proceedings, 91 \\
http://ivscc.gsfc.nasa.gov/publications/gm2012/oyama.pdf


\bibitem[Povich (2009)]{povich09}
Povich, M. S. 2009, Ph.D. Thesis,
The University of Wisconsin - Madison

\bibitem[Povich \& Whitney(2010)]{povich10}
Povich, M. S., \& Whitney, B. A. 2010, 
\apjl, 714, L285

\bibitem[Povich et al.(2016)]{povich16}
Povich, M. S., Townsley, L. K., Robitaille, T. P., Broos, P. S., Orbin, W. T., King, R. R., Naylor, T., \& Whitney, B. A. 2016, 
\apj, 825, 125

\bibitem[Ragan et al.(2014)]{ragan14}
Ragan, S. E., Henning, Th., Tackenberg, J., Beuther, H., Johnston, K. G., Kainulainen, J., \& Linz, H. 2014,
\aap, 568, A73

\bibitem[Rathborne et al.(2009)]{rathborne09}
Rathborne, J. M., Johnson, A. M., Jackson, J. M., Shah, R. Y., \& Simon, R.  2009, \apjs, 182, 131

\bibitem[Reid et al.(2014)]{reid14}
Reid, M. J. et al. 2014,
\apj, 783, 130

\bibitem[Reid et al.(2016)]{reid16}
Reid, M. J., Dame, T. M., Menten, K. M., \& Brunthaler, A. 2016,
\apj, 823, 77

\bibitem[Rigby et al.(2016)]{rigby16}	
Rigby, A. J. et al. 2016, 
\mnras, 456, 2885

\bibitem[Rosolowsky et al.(2008)]{rosolowsky08}
Rosolowsky, E. W., Pineda, J. E., Kauffmann, J., \& Goodman, A. A. 2008, 
\apj, 679, 1338

\bibitem[Sanders et al.(1984)]{sanders84}
Sanders, D.B., Solomon, P.M., \& Scoville, N.Z. 1984, \apj, 276, 182

\bibitem[Sawada et al.(2008)]{sawada08}
Sawada , T. et al. 2008, \pasj, 60, 445

\bibitem[Schloerb, Snell and Schwartz(1987)]{schloerb87}
Schloerb, F. P., Snell, R. L., \& Schwartz, P. R. 1987,
\apj, 319, 426

\bibitem[Schuller et al.(2009)]{schuller09}
Schuller, F. et al. 2009, 
\aap, 504, 415

\bibitem[Scoville(2013)]{scoville13}
Scoville, N. Z. 2013, 
Secular Evolution of Galaxies, ed. J. Falc\'on-Barroso \& J. H. Knapen (Cambridge University Pres), 491

\bibitem[Scoville \& Solomon(1974)]{scoville74}
Scoville, N. Z., Solomon, P. M. 1974, 
\apjl, 187, 67

\bibitem[Sousbie(2011)]{sousbie11}
Sousbie, T. 2011, \mnras,  414, 350

\bibitem[Tackenberg et al.(2013)]{tackenberg13}
Tackenberg, J., Beuther, H., Plume, R., Henning, T., Stil, J., Walmsley, M., Schuller, F., \& Schmiedeke, A. 2013,
\aap, 550, A116

\bibitem[T\'oth et al.(2014)]{toth14}
T\'oth , L. V. et al. 2014, 
\pasj, 66,17

\bibitem[Watson et al.(2008)]{watson08}
Watson, C. et al. 2008,
\apj, 681, 1341

\bibitem[Williams et al.(1994)]{williams94}
Williams, J. P., de Geus, E. J., \& Blitz, L. 1994, \apj, 428, 693

\bibitem[Xu et al.(2011)]{xu11}
Xu, Y., Moscadelli, L., Reid, M. J., Menten, K. M., Zhang, B., Zheng, X. W., \& Brunthaler, A. 2011,
\apj, 733, 25

\bibitem[Yamagishi et al.(2016)]{yamagishi16}
Yamagishi, M. et al. 2016,
\apj, 833, 163
	

\end{thebibliography}
\end{document}